\documentclass[sigplan,10pt,nonacm]{acmart}
\usepackage{amsmath}
\usepackage{booktabs}
\usepackage{tabularx}
\usepackage{graphicx}
\usepackage{placeins}
\usepackage{stfloats}
\usepackage{xcolor}
\usepackage{url}
\definecolor{logicalblue}{HTML}{0055CC}
\definecolor{qecteal}{HTML}{007A78}
\definecolor{supplyorange}{HTML}{B86700}
\newcommand{\sys}{FTWeave}
\newcommand{\sysfull}{\textbf{\sys{}}}
\newcommand{\us}{\ensuremath{\,\mu\mathrm{s}}}
\newcommand{\ProtectionComparisonFigure}{%
\begin{figure*}[!tbp]
\centering
\includegraphics[width=.96\textwidth]{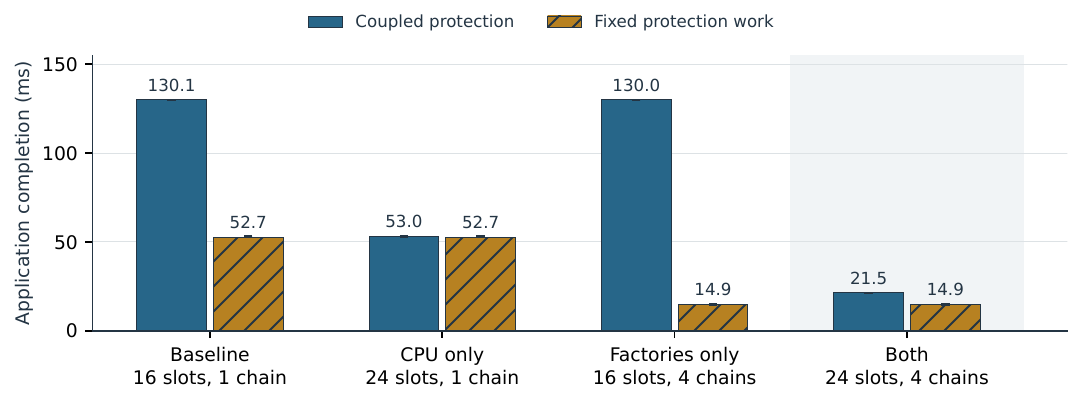}
\caption{Multiplier CPU-only, factory-only, and joint upgrades from 16 CPU slots
and one chain to 24 slots and/or four chains. Fixed work favors factories
alone, while coupled protection needs both upgrades for $6.06\times$
speedup. Bars show three-seed completion medians and ranges.
Section~\ref{sec:study-scope} defines both models.}
\label{fig:protection-comparison}
\Description{The baseline, CPU-only, factories-only, and joint cases take
130.1, 53.0, 130.0, and 21.5 milliseconds under coupled protection.
Fixed work predicts 52.7, 52.7, 14.9, and 14.9 milliseconds. The same
four configurations therefore imply different upgrade sequences.}
\end{figure*}}

\newcommand{\ApplicationProvisioningFigure}{%
\begin{figure*}[!tbp]
\centering
\includegraphics[width=.96\textwidth]{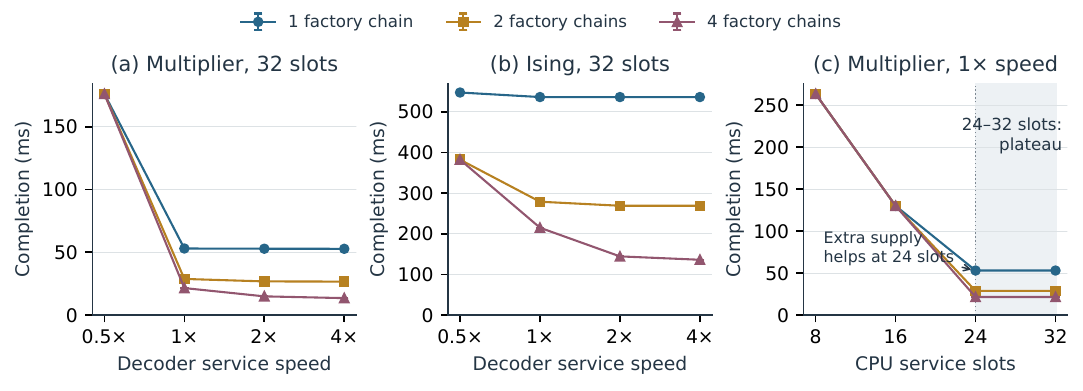}
\caption{Decoder latency, decoder capacity, and factory supply constrain
 different parts of execution. (a,b) At 32 CPU slots, faster decoding
helps most when additional factories supply states. (c) At original
 decoder speed, more factories become useful for the multiplier at
24 CPU slots and then plateau through 32 slots, marked by shading.
Each point is a three-seed median, with observed ranges.
Lines connect tested settings. Panels use separate completion-time scales.}
\label{fig:application-provisioning-matrix}
\Description{Three curves show one, two, and four factory chains.
The multiplier and Ising speed sweeps flatten with one chain. The multiplier
capacity sweep shows overlapping curves at eight and sixteen slots, then
separate curves at twenty-four and thirty-two slots.}
\end{figure*}}

\newcommand{\OwnershipInteractionFigure}{%
\begin{figure}[!htbp]
\centering
\includegraphics[width=\columnwidth]{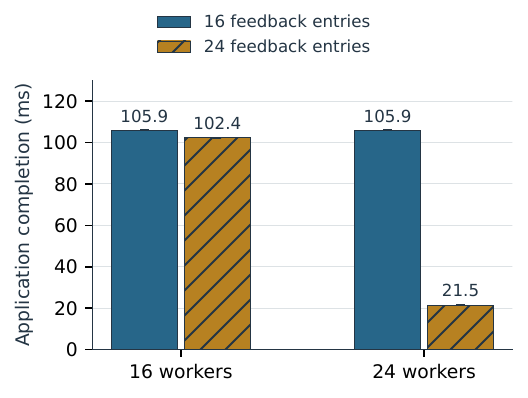}
\caption{Multiplier workers and feedback entries must grow together for
$4.93\times$ speedup. All cases use four chains and end-to-end admission.
Bars show three-seed completion medians and ranges.}
\label{fig:ownership-interaction}
\Description{With sixteen feedback entries, sixteen and twenty-four workers
both take 105.9 milliseconds. With twenty-four entries, they take 102.4
and 21.5 milliseconds. Storage and worker capacity are independent limits.}
\end{figure}}

\newcommand{\OwnershipTailFigure}{%
\begin{figure*}[!t]
\centering
\includegraphics[width=.96\textwidth]{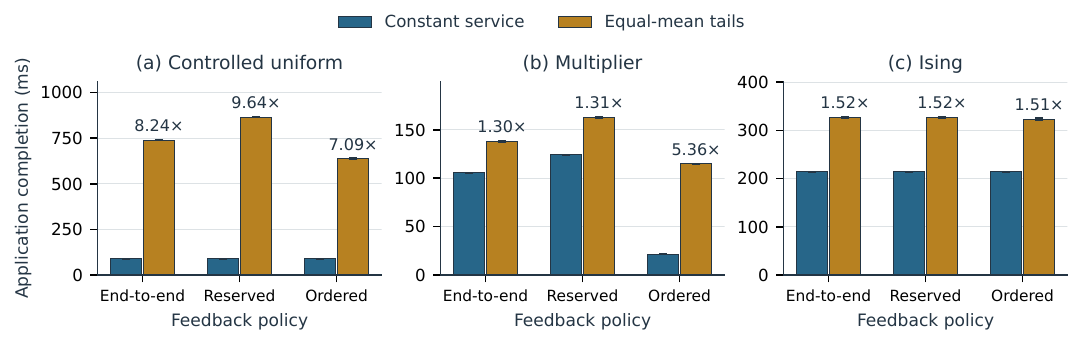}
\caption{Ownership sensitivity to equal-mean service tails with 24 workers,
sixteen feedback entries, four chains, and the same explicit classical
path. Bars show three-seed medians and ranges. Labels give median paired
tail/constant ratios. Ordered publication remains fastest for the
multiplier. Panels use separate time scales.}
\label{fig:ownership-tails}
\Description{Three panels compare constant and tail-heavy service under
end-to-end credits, reserved feedback, and ordered publication. Controlled
completion rises from about 90 milliseconds to 739, 865, and 636 milliseconds.
The multiplier takes 138, 163, and 115 milliseconds with tails, while Ising
takes about 323 to 327 milliseconds.}
\end{figure*}}

\newcommand{\ControllerCapacityFigure}{%
\begin{figure*}[!tbp]
\centering
\includegraphics[width=.96\textwidth]{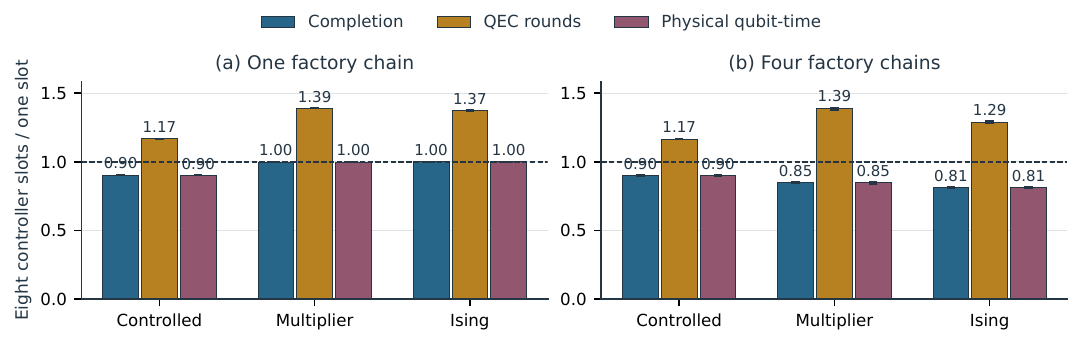}
\caption{Controller capacity affects progress and protection issuance. Bars show
the three-seed median and range of matched eight-slot/one-slot ratios at
24 CPU slots. The dashed line is unchanged behavior. With four chains,
more controller capacity reduces completion and physical qubit-time
despite more executed rounds.}
\label{fig:controller-capacity}
\Description{Two panels compare one and four factory chains. With four
chains, multiplier and Ising completion fall to 0.85 and 0.81 times their
one-controller-slot values, while QEC rounds rise to 1.39 and 1.29 times.
Physical qubit-time falls with completion. With one chain, application
completion changes little despite increased round counts.}
\end{figure*}}

\newcommand{\ApplicationWindowFigure}{%
\begin{figure*}[!tbp]
\centering
\includegraphics[width=.96\textwidth]{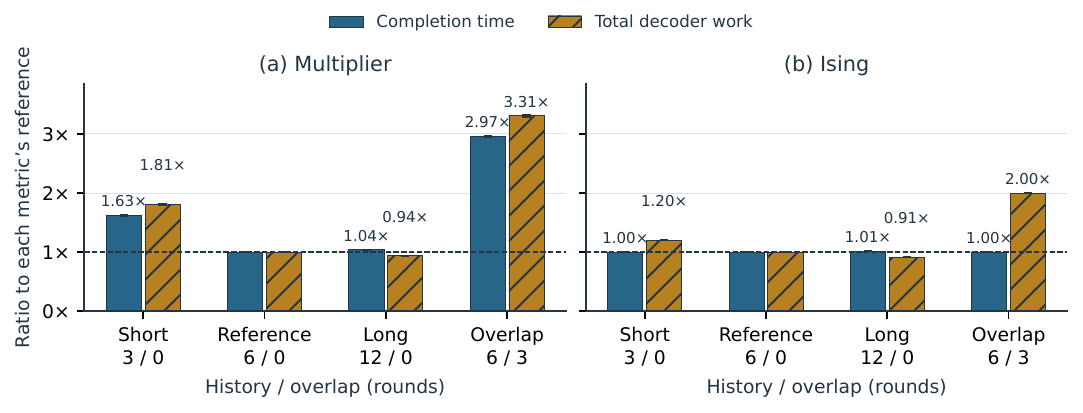}
\caption{Less decoder work need not mean earlier completion. Blue bars show
completion and gold bars summed decoder service, each normalized to its
six-round, zero-overlap reference. Ratios are paired by seed and show
medians and ranges. Both applications use one chain, 32 CPU slots,
and 24 outstanding windows.}
\label{fig:application-window-work}
\Description{For each application, paired bars compare completion and decoder
work at histories of three, six, and twelve rounds and at six rounds with
three rounds of overlap. Multiplier overlap gives 2.97 times completion and
3.31 times work. Ising overlap doubles work with nearly unchanged completion.}
\end{figure*}}

\newcommand{\ApplicationVariabilityFigure}{%
\begin{figure*}[!tbp]
\centering
\includegraphics[width=.96\textwidth]{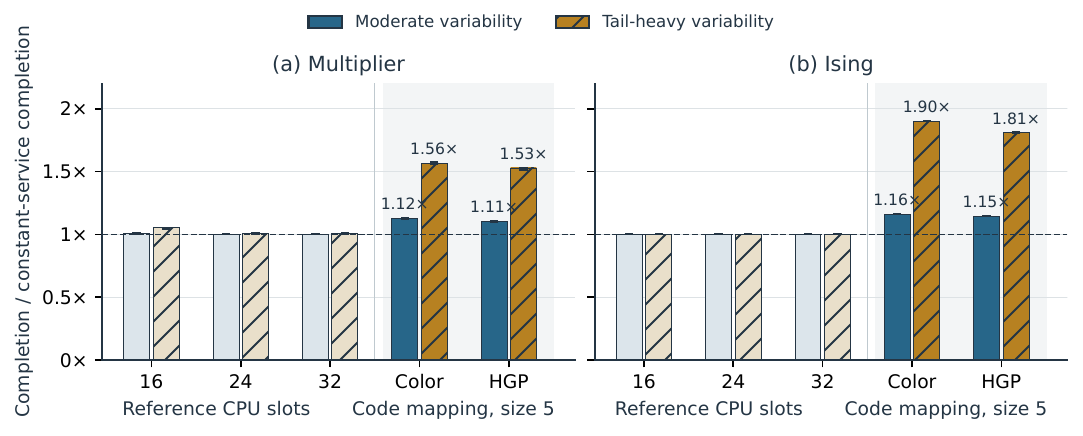}
\caption{Completion relative to the matched constant-service case, paired by seed,
with three-seed medians and ranges. Reference controls use the labeled CPU
capacities and 24 outstanding windows. Color/HGP use size coordinate five, 256 CPU
slots, and 256 windows. All use one chain. The shared scale makes
$1.90\times$ equal to 90\% longer completion.}
\label{fig:application-tail-mappings}
\Description{Reference-mapping bars stay close to one. Tail-heavy service
gives 1.56 and 1.53 times completion for the multiplier under color and HGP,
and 1.90 and 1.81 times for Ising. Moderate variability is shown alongside
each tail-heavy case. All comparisons preserve the expected service mean.}
\end{figure*}}

\newcommand{\ApplicationFamilyFigure}{%
\begin{figure*}[!tbp]
\centering
\includegraphics[width=.94\textwidth]{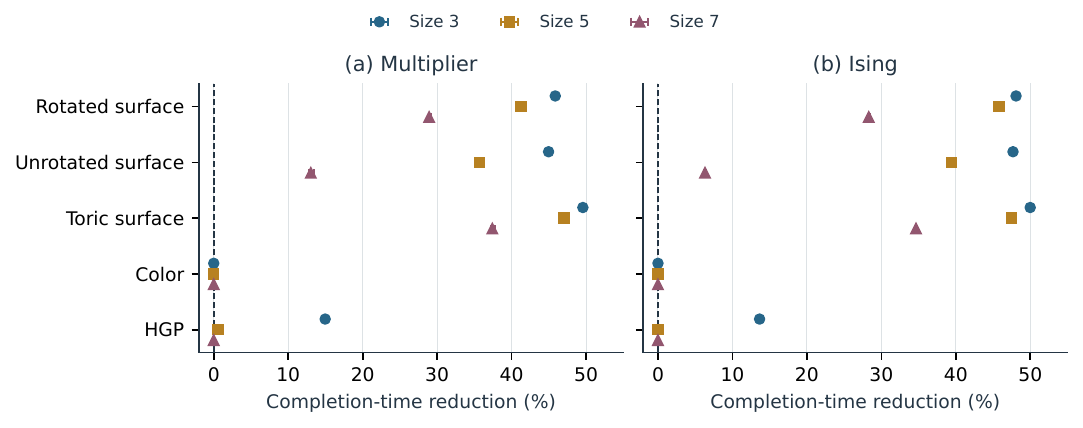}
\caption{Second-chain completion reduction, $100(1-T_2/T_1)$, paired by seed.
Here $T_1$ and $T_2$ are one- and two-chain completion times.
Symbols identify sizes and whiskers show three-seed ranges. Each pair uses
256 CPU slots, 256 outstanding windows, and unchanged code geometry,
service, and recipes. Absolute times appear in
Figure~\ref{fig:application-family-absolute}.}
\label{fig:application-family-matrix}
\Description{A second chain helps surface and toric mappings most.
Color and several HGP configurations gain little. Each point compares
supply within one mapping, with no cross-family reliability equivalence.}
\end{figure*}}

\begin{document}
\title[FTWeave]{FTWeave: Composing Logical Execution, Real-Time Decoding, and Magic-State Supply in Tightly Integrated Quantum--Classical Systems}

\author{Panayiotis Christou}
\affiliation{%
  \institution{Department of Computer Science, Fordham University}
  \city{New York}
  \country{USA}
}
\email{pc33@fordham.edu}

\author{Shuwen Kan}
\affiliation{%
  \institution{Department of Computer Science, Fordham University}
  \city{New York}
  \country{USA}
}
\email{sk107@fordham.edu}

\author{Ying Mao}
\affiliation{%
  \institution{Department of Computer Science, Fordham University}
  \city{New York}
  \country{USA}
}
\email{ymao41@fordham.edu}

\begin{abstract}
Fault-tolerant quantum applications depend on both decoded measurement
information and a continuing supply of magic states. While an operation
waits for either, its logical data require protection that produces more
syndrome data and decoding work. Architects must provision the quantum
and classical parts of the machine together, before complete systems
are available to measure.

\sysfull{} is a system-level simulator for this architectural exploration.
It accepts OpenQASM~2/3 circuits, CUDA-Q Logical compiler artifacts,
Stim circuits, and provider-derived physical inputs through adapters
that preserve operation and
measurement dependencies. Users bind these inputs to quantum resources,
classical processors, memory, communication, and factory configurations.
One event timeline determines application progress, resource ownership,
and continuing protection. This supplies the system model needed to
connect circuit and decoder evidence to application performance.

Controlled demand studies and complete multiplier and Ising circuits
show why the coupling changes provisioning decisions. A fixed-work
approximation predicts a $3.55\times$ multiplier speedup from additional
factories where the coupled model gives almost none. Explicit ownership
also changes useful capacity: increasing decoder workers alone has no
benefit in a storage-constrained case, while increasing workers and
feedback entries together gives $4.93\times$ speedup. Selected color and hypergraph-product code mappings take 53--90\%
longer under decoder service variability, despite an unchanged mean.
These comparisons identify useful resource combinations and distinguish
protection work from the elapsed lifetime of quantum data.
\end{abstract}
\maketitle
\section{Introduction}
\label{sec:intro}

A fault-tolerant quantum operation can be ready in the program and still
wait for decoded measurement information or a suitable magic state.
Its logical qubits remain protected throughout that wait. Repeated
quantum error correction (QEC) produces syndrome data and decoding work, which compete with
factory control and later operations for classical resources. An
architectural delay therefore changes how much work the machine must
perform as well as when the application finishes.

Consider a T operation whose required measurements have been decoded
but whose factory output is unavailable. Faster decoding cannot release
that operation. A second factory may shorten its wait, allow the program
to finish earlier, and avoid protection work that would otherwise be
necessary. Once state supply is sufficient, further factories may have
no benefit. The useful upgrade depends on the application's current
information, material, and resource dependencies.

Architects must make these decisions before complete, tightly integrated
fault-tolerant machines are available. Component measurements establish
decoder service or factory outcomes at particular operating points.
Circuit compilation supplies the operations and their dependencies.
Predicting application completion also requires knowing when those
operations become reachable, which resources they hold while waiting,
and how much protection their waits generate. A system-level simulator
can combine these inputs and test resource combinations while retaining
the distinction between measured costs and architectural assumptions.

Stim simulates stabilizer circuits and produces samples and detector
error models~\cite{stim2021}. PyMatching decodes syndrome data using
minimum-weight perfect matching~\cite{higgott2025}. CUDA-Q Logical
compiles logical programs through QEC realizations to physical schedules
and derives resource estimates from those compiler
artifacts~\cite{cudaqlogical2026}. These tools provide circuit behavior,
decoder implementations, and compiled structure. A system study must
also determine how their work competes for finite resources over time.
In particular, waiting for a decoder or a factory can prolong quantum
protection and create work that a fixed circuit or service-cost total
does not capture.

Building that study from component libraries requires a connecting model:
when measurements become available, how requests acquire and release
processors and storage, when feedback permits progress, and how factories
replenish finite inventory. \sysfull{} supplies these execution rules in one
simulator. An architect can bind a compiled workload and component
profiles to a proposed machine, then vary its capacities and policies
without implementing a separate scheduler and resource model for each
study. The resulting question is concrete: can the application use
additional factory or decoder capacity, and how much continuing
protection does the resulting change in progress create or avoid?

The input interfaces connect existing workflows to this model.
OpenQASM~2/3 and CUDA-Q Logical supply logical or lowered physical work,
Stim supplies physical measurement dependencies, and provider bindings
retain native operations, movement, and control decisions. Adapters bind
that structure to recipes, profiles, and resources for application and
physical-execution studies.

\paragraph{Contributions.}
\begin{itemize}
  \item \textbf{A coupled execution model.}
  Reachable operations, their information requirements, and finite
  state supply determine how long data remain protected. The model
  captures the additional QEC work generated during a wait and its
  effect on subsequent progress.
  \item \textbf{Explicit ownership across the quantum--classical system.}
  The model separates admission and release for quantum processing unit
  (QPU) allocations, spatial claims, processors, memory, copies, and
  provider state. These lifetimes expose contention that persists after
  computation finishes and determine when another operation can use
  the same capacity.
  \item \textbf{Reusable input interfaces.}
  Adapters import OpenQASM~2/3, CUDA-Q Logical, Stim, and provider inputs.
  Each binding retains source operations and dependencies and supplies
  timing and resource parameters. These interfaces connect existing
  application and compiler workflows to system studies using one
  execution model.
  \item \textbf{Architectural comparisons that explain useful capacity.}
  Controlled demand isolates individual mechanisms, and complete
  multiplier and FTCircuitBench Ising circuits test their interaction.
  Matched interventions show when fixed protection work mispredicts a
  useful upgrade, when workers need more feedback storage, and how
  controller capacity, code mappings, windows, and service tails change
  provisioning decisions.
\end{itemize}

For the multiplier, fixed protection work predicts a $3.55\times$
factory-upgrade benefit at 16 CPU slots where coupled execution gives
almost none. Coupled execution reaches $6.06\times$ baseline speed
with 24 CPU slots and four factory chains. A separate ownership study finds that
extra workers help only when feedback storage grows with them.

Sections~\ref{sec:design}--\ref{sec:implementation} define and validate the
model and adapters. Sections~\ref{sec:application-method}--\ref{sec:application-results}
then connect the experimental method to controlled and complete-application studies.

\section{Background and Related Work}
\label{sec:background}

\subsection{Logical Computation and Classical Readiness}

Fault tolerance encodes logical qubits across physical qubits and
repeatedly measures error syndromes. The parity checks reveal errors
without directly reading the encoded data. A decoder infers a recovery. Logical failure occurs when the physical
error and chosen recovery together implement a nontrivial logical
operator on the code space. Code family,
distance, extraction circuit, and noise determine the relevant timing
and error characteristics \cite{fowler2012surface,google2025quantum,terhal2015}.

Pauli corrections can be tracked in a software frame and propagated
through Clifford operations. Logical work and extraction may therefore
continue while decoding lags. Adaptive measurements and non-Clifford
feed-forward can require resolved information \cite{battistel2023}.
A decoder needs enough throughput to keep up with sustained
syndrome production. The program's synchronization points determine
when an individual decoding delay prevents logical progress.

Windows trade history-collection delay for service efficiency and
concurrency \cite{skoric2023parallel}. Sparse blossom improves minimum-weight perfect matching (MWPM)
throughput \cite{higgott2025}. AFS studies decoder hardware and shared
storage and bandwidth \cite{das2022afs}. Reaction-time models connect
classical service and communication to logical execution \cite{khalid2026}.
These costs affect when information becomes usable, which can differ
from the instant a decoder returns a result.

\subsection{Magic States as a Supply Process}

Distillation consumes noisy states and postselects higher-quality T or
CCZ resources \cite{bravyi2005,bravyi2012magic}. Mapping and scheduling
shape factory space--time cost \cite{ding2018magic,litinski2019}.
Cultivation and catalyzed conversion provide other preparation paths
\cite{gidney2024magic,gidney2019}. A catalyst remains a reusable resource
with its own ownership, separate from consumed inputs and new outputs.

A \emph{production chain} connects preparation stages and stores to a
usable output. Each chain in our evaluation contains raw-T preparation
and 15-to-1 distillation. Replication adds quantum resources while sharing
classical capacity and final inventory. Attempt latency, launch cadence,
concurrency, rejection, and storage determine when a consumer receives
a state. Equal T counts can exert different pressure when dependencies
group or spread consumption. Compiled representations such as
FTCircuitBench retain the circuit structure needed to expose this
variation \cite{ftcircuitbench2026}.

\subsection{What Joint System Modeling Adds}
\label{sec:prior_systems}

\emph{Price and Payoff} couples circuit scheduling with stochastic magic-state
production and studies how nondeterminism changes useful factory capacity
\cite{awasthi2026price}. Its supply-driven scheduling question is closely
related to ours. \sys{} additionally follows the decoding work generated
while protected data wait, allowing a supply intervention to change
classical demand as well as the consumption schedule.

\emph{Stalls and Spequlation} decomposes logical operations into control,
execution, and decoding, and studies speculative overlap
\cite{awasthi2026stalls}. This exposes parallelism hidden by atomic logical
operations. \sys{} follows fixed information requirements and finite
material ownership through such execution boundaries. The studies here
use those dependencies to assess provisioning without speculative rollback.

Elastic QEC decoders treat decoding as a shared accelerator service and
study capacity planning and scheduling under fluctuating demand
\cite{maurya2026elastic}. Our comparisons ask how the useful classical
capacity changes when factory supply and application progress also
change the duration of protection. They report both completion and the
protection work produced along the way.

HybridQC models typed hybrid stages and contention
\cite{hybridqc2026}. StreamingQEC follows syndrome movement, decoding,
and feedback within supplied protected intervals \cite{streamingqec2026}.
\sys{} makes the lifetime of those intervals an outcome of reachable
logical work, required information, and available states. These models
share a finite-resource simulation approach, also used in architectural
tools such as ns-3 and gem5 \cite{ns3,binkert2011gem5}.

Stim, PyMatching, and CUDA-Q Logical supply circuit behavior, decoding,
and compiled logical or physical structure~\cite{stim2021,higgott2025,cudaqlogical2026}.
\sys{} binds this evidence to shared resources, finite state supply,
and protection that continues while the application waits. CUDA-Q,
cuQuantum, and CUDA-Q QEC also supply execution and component-analysis
tools~\cite{cudaq,cuquantum2023,cudaqqec}.
Other studies address code choice~\cite{hetec2025}, trapped-ion protection
and selective compilation~\cite{jones2026,iswitch2026}, quantum low-density
parity-check (qLDPC) layout and distillation~\cite{acqc2026}, cryogenic
decoding~\cite{ueno2021qecool}, communication~\cite{rao2026}, and architecture
representations~\cite{tierx2026,rteaal2026}. These inputs help determine
which resource change releases a limiting dependency and how it changes
continuing protection.

\section{System Design and Modeling}
\label{sec:design}

A logical operation advances when its required information and input states
are ready and its execution resources can be acquired. Its data remain
protected while it waits.

\begin{figure*}[t]
\centering
\includegraphics[width=\textwidth]{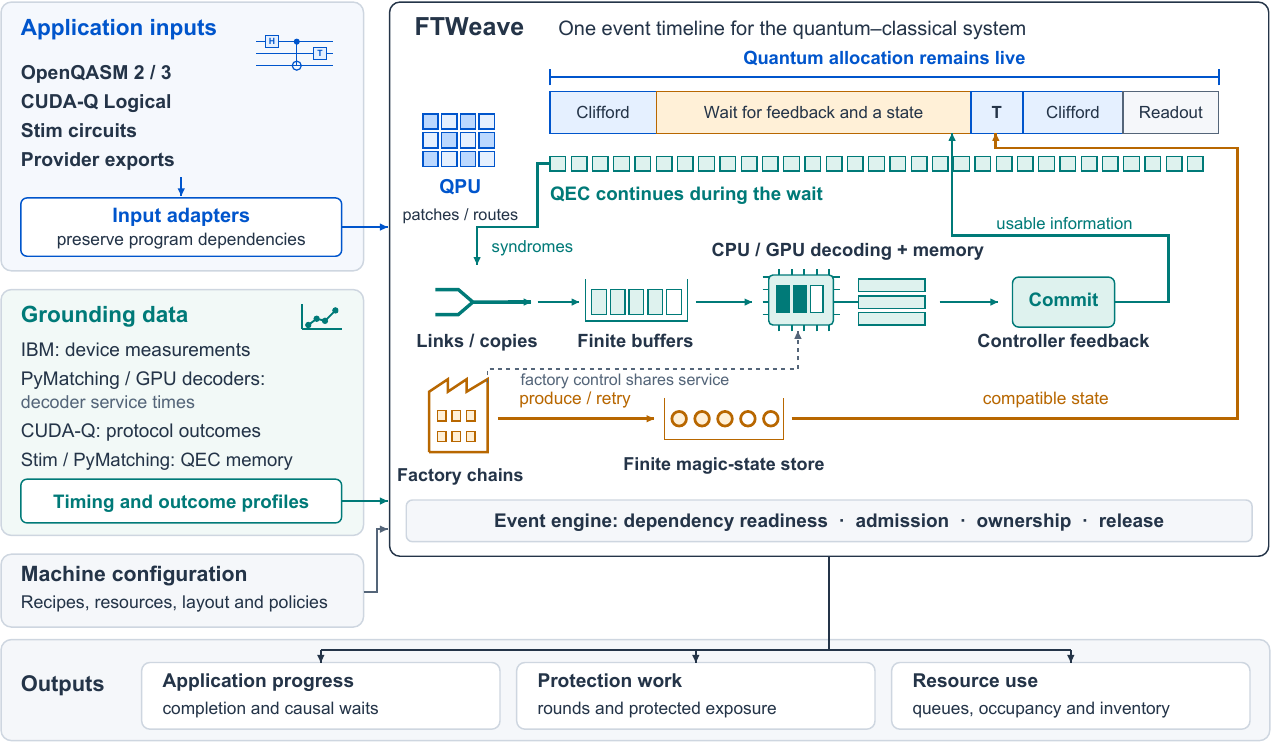}
\caption{\sys{} combines dependency-preserving inputs, grounding profiles, and
machine settings on one event timeline. The waiting T operation needs
usable feedback and a compatible magic state while continuing QEC and
factory control share classical resources. Outputs record progress,
protection work, and resource use. Timelines and counts are schematic.}
\label{fig:architecture}
\Description{Three input panels feed FTWeave. Application inputs include OpenQASM, CUDA-Q Logical, Stim circuits, and provider exports, connected through dependency-preserving adapters. Grounding data include IBM device measurements, PyMatching and GPU decoder service times, CUDA-Q protocol outcomes, and Stim with PyMatching QEC memory characterization. A separate machine configuration specifies recipes, resources, layouts, and policies. Within FTWeave, a quantum allocation contains Clifford work, a waiting T operation, and readout. QEC rounds continue through the wait and feed links, copies, finite buffers, processors, memory, and controller feedback. Factory chains share classical service and replenish a finite magic-state store. Feedback and a compatible state enable the T operation. One event engine manages readiness, admission, ownership, and release. An enclosing Outputs panel contains application progress, protection work, and resource use.}
\end{figure*}

Figure~\ref{fig:architecture} follows a waiting T operation. Adapters supply
dependencies, profiles supply service and outcomes, and machine settings
bind work to recipes and resources. Continuing QEC and factory control
compete for shared classical service, while factories replenish finite
inventory. One event engine records how these dependencies determine
progress, protection work, and resource use.

\subsection{Application Demand and Logical Recipes}
\label{sec:demand}

The demand graph contains logical operations, Clifford intervals,
measurements, branches, and joins. Each operation identifies its logical
support and predecessors. A recipe adds physical duration, resource
requirements, information dependencies, and typed material consumption.
Compilation fixes the operations and dependencies of a static circuit.
For adaptive inputs, the graph also retains branches and their measured
conditions. Execution follows the reached branch. Contention determines
when those operations can advance.

Logical width, physical footprint, and spatial occupancy describe
different constraints. Width counts the logical data carried by the
application. The code mapping determines the physical data and ancilla
capacity needed to protect those data. A layout additionally determines
where patches reside and which operations compete for routing space.
Two independent T operations can therefore have both states and feedback
ready yet wait for the same injection interface. Adding factories cannot
remove that spatial conflict. The waiting data continue to require QEC.

An optional fixed layout binds data and magic patches, oriented boundaries,
ancillas, and connected routes to exclusive tile resources. For this
binding, an exclusive quantum lease preserves resident data and magic
patches between gates. Operations within that lease acquire tile claims
inside the already allocated device. They do not allocate the device a
second time. The lease excludes other allocations, while tile claims
control concurrency among its own operations.
A spatial operation acquires its patch and route claims together and
holds them through completion. A contiguous merge, measurement, and split
can share that reservation. Their durations partition the recipe's
existing service interval, so spatial binding introduces contention
without charging the operation's duration twice. Reuse is possible only
after the owning operation releases its claims.

The logical and physical paths supply different timing information.
A logical recipe specifies an operation's service and information needs.
A physical input retains individual operations, measurements, and their
dependencies, with explicit gate, readout, reset, and shared-resource
bindings where supplied. Its extraction schedule determines when the
measurements needed by a detector become available. Decoder input cannot
arrive before that boundary. Both paths feed the same resource and
information owners. A compiler dependency edge by itself supplies
neither a hardware duration nor a decoding result.

For example, an extraction measurement becomes visible after its readout
operation. Its detector can then feed decoding while the measured ancilla
resets for reuse. A correction depending on that detector waits for the
selected feedback frontier. An operation using the ancilla instead waits
for reset and any remaining spatial claims. The data allocation stays
protected through both paths. These distinct dependencies let a physical
input represent feedback overlapping reuse without treating measurement
completion as permission for every successor to start.

The controlled generator varies grouping and spacing independently of
total demand. For epoch $k$, let $g_k$ be preceding Clifford work in
protection rounds and $b_{k,m}$ the consumptions of material class $m$.
Matched workloads retain
\begin{equation}
 \sum_k g_k=R,\qquad \sum_k b_{k,m}=N_m,
 \label{eq:matched-demand}
\end{equation}
including prefix and tail work in $R$. Logical allocation, recipes,
circuits, and shots also remain fixed. At round interval $\tau$, the
requested Clifford time is $R\tau$. Synchronization and resource waits
determine the additional elapsed time.

\emph{Uniform} demand uses evenly spaced consumers.
\emph{Burst} demand groups $b>1$ consumers into $K=N/b$ epochs with intervals
$g_k\approx R/K$. \emph{Clustered} demand places individual consumers
in high- and low-density phases. If fraction $f$ of the demand has density
$c$ times the remainder, their work budgets are
\begin{equation}
 R_H=\frac{fR}{c(1-f)+f},\qquad R_L=R-R_H.
 \label{eq:phase-budgets}
\end{equation}
Thus $f=1/2,c=4$ place half the demand in one fifth of the work.
\emph{Clustered-burst} combines the phase budgets with grouped consumers.
Integer apportionment preserves exact totals. Independent consumers use
disjoint logical supports and acquire separate states before joining. An atomic consumer requires its whole state batch together.
Appendix~\ref{app:demand-policy} defines stochastic spacing and persistent
variants. Paired interventions reuse the saved demand graph, with separate
seeds for demand generation and runtime outcomes.

\subsection{Continuous Protection and Fixed Information Requirements}
\label{sec:frontiers}
\label{sec:windows}

Waiting extends protected lifetime. A live logical allocation requests
syndrome work at its configured cadence, subject to the resources and
backpressure of its extraction path. This continues during logical,
material, and execution-resource waits. When the protected computation
finishes, generation stops. Issued work drains before the allocation is
released.

For a protected source, $P(t)$ and $D(t)$ denote the last consecutively
produced and decoded rounds. Decode backlog is $B(t)=P(t)-D(t)$.
An out-of-order result advances $D$ only after earlier gaps are resolved.
Extraction, transfer, provider admission, decoding, controller application,
and ordered commit have distinct completion points. The consumer selects
the point at which information is usable for its recipe.

A reached barrier captures its requirement once. For a barrier reached
at $a_k$ that needs the current history of one source,
\begin{equation}
 r_k=P(a_k),\qquad
 t_k^{\mathrm{QEC}}=\inf\{t\ge a_k:F_k(t)\ge r_k\},
 \label{eq:barrier}
\end{equation}
where $F_k$ is its selected decoded, applied, or committed frontier.
A barrier with several sources retains the corresponding requirements
for each. Later protection rounds add work without moving these targets.
Otherwise a waiting operation could continually demand information that
its own wait had just caused the machine to generate.

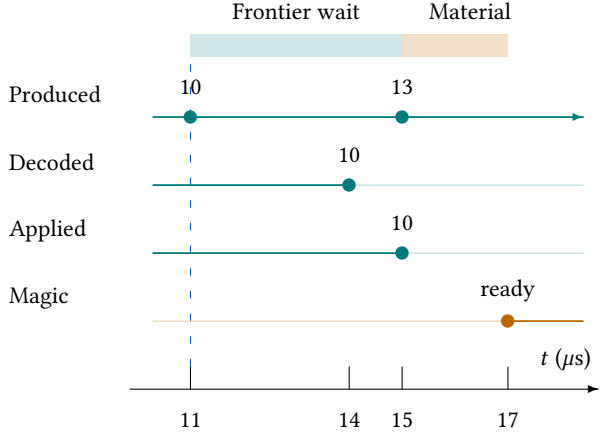
\begin{figure}[t]
\centering
\setlength{\unitlength}{1mm}
\begin{picture}(83,58)
% Colored wait intervals share the physical-time scale below.
\put(27,48){\color{qecteal!18}\rule{28mm}{3mm}}
\put(55,48){\color{supplyorange!20}\rule{14mm}{3mm}}
\put(41,54){\makebox(0,0){\small Frontier wait}}
\put(64,54){\makebox(0,0){\small Material}}
\multiput(27,7)(0,3){14}{\color{logicalblue}\line(0,1){1}}
\put(3,43){\makebox(0,0)[l]{\small Produced}}
\put(3,34){\makebox(0,0)[l]{\small Decoded}}
\put(3,25){\makebox(0,0)[l]{\small Applied}}
\put(3,16){\makebox(0,0)[l]{\small Magic}}
\put(22,40){\color{qecteal!30}\line(1,0){57}}
\put(22,31){\color{qecteal!30}\line(1,0){57}}
\put(22,22){\color{qecteal!30}\line(1,0){57}}
\put(22,13){\color{supplyorange!35}\line(1,0){57}}
\linethickness{0.65pt}
\put(22,40){\color{qecteal}\vector(1,0){57}}
\put(22,31){\color{qecteal}\line(1,0){26}}
\put(22,22){\color{qecteal}\line(1,0){33}}
\put(27,40){\color{qecteal}\circle*{1.6}}
\put(55,40){\color{qecteal}\circle*{1.6}}
\put(48,31){\color{qecteal}\circle*{1.6}}
\put(55,22){\color{qecteal}\circle*{1.6}}
\put(27,44){\makebox(0,0){\small 10}}
\put(55,44){\makebox(0,0){\small 13}}
\put(48,35){\makebox(0,0){\small 10}}
\put(55,26){\makebox(0,0){\small 10}}
\put(69,13){\color{supplyorange}\circle*{1.8}}
\put(69,13){\color{supplyorange}\line(1,0){10}}
\put(69,17){\makebox(0,0){\small ready}}
\linethickness{0.4pt}
\put(19,4){\vector(1,0){62}}
\put(27,4){\line(0,1){3}}
\put(48,4){\line(0,1){3}}
\put(55,4){\line(0,1){3}}
\put(69,4){\line(0,1){3}}
\put(27,0){\makebox(0,0){\small 11}}
\put(48,0){\makebox(0,0){\small 14}}
\put(55,0){\makebox(0,0){\small 15}}
\put(69,0){\makebox(0,0){\small 17}}
\put(80,8){\makebox(0,0)[r]{\small $t$ ($\mu$s)}}
\end{picture}
\caption{Captured-frontier example. The dashed line fixes round 10 at
$11\us$. Teal markers show QEC progress. The applied target resolves at
$15\us$, and the orange batch becomes ready at $17\us$. Protection
continues during both waits. Resources are available at $17\us$.}
\label{fig:frontier}
\Description{A blue dashed line marks capture of round 10 at 11 microseconds. Teal QEC markers show decoding at 14 and application at 15 microseconds. The produced frontier has reached round 13 by then. An orange marker shows material ready at 17 microseconds. Shaded intervals show the frontier wait and the subsequent material wait in this example.}
\end{figure}

In Figure~\ref{fig:frontier}, capture at $11\us$ fixes round 10.
Decoding reaches it at $14\us$ and application at $15\us$, even though
production has reached round 13. Material arrives at $17\us$. The
operation can then proceed if its resource bundle is admissible.

A decoding window becomes ready when its declared history exists.
History length, overlap, newly resolved range, outstanding-window limit,
and incomplete-tail rule are explicit inputs. The evaluated fixed-stride paths include a
partial final window when protection stops, resolving only the remaining
rounds. They retain the configured call service and payload conventions
for that final window. An overlapping window processes old history again,
but advances progress only over its new range. A window follows an acyclic
service graph, possibly through a predecoder and conditional fallback.
Different windows can occupy different stages concurrently. Increasing
worker or window capacity admits more concurrent work. It does not divide
a call's service duration. A waiting deadline records a budget violation. Any cancellation, timeout outcome, or recovery must follow its declared
policy rather than silently shortening a successful service call.

\subsection{Finite Factories and State Inventory}
\label{sec:factories}
\label{sec:knobs}

A factory's nominal rate does not determine when a consumer receives a
state. Raw supply, distillation stages, verification, acceptance, storage,
and transfers must complete. Each deployed factory binds these stages
to resources and input/output stores. Lots retain class, quantity,
quality, readiness, location, and provenance, allowing consumers to
request compatible states at their required quality.

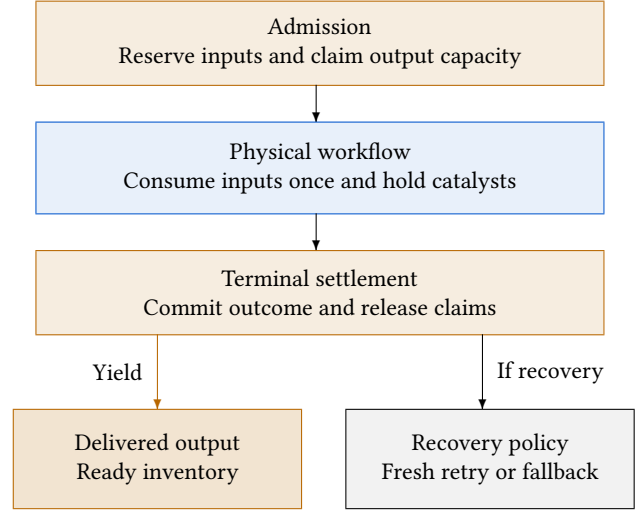
\begin{figure}[t]
\centering
\setlength{\unitlength}{1mm}
\begin{picture}(83,71)
\put(4,57){\color{supplyorange!12}\rule{75mm}{11mm}}
\put(4,57){\color{supplyorange}\framebox(75,11){}}
\put(4,57){\makebox(75,11){\parbox{71mm}{\centering\small
Admission\\Reserve inputs and claim output capacity}}}
\put(41,57){\vector(0,-1){5}}
\put(4,40){\color{logicalblue!9}\rule{75mm}{12mm}}
\put(4,40){\color{logicalblue}\framebox(75,12){}}
\put(4,40){\makebox(75,12){\parbox{71mm}{\centering\small
Physical workflow\\Consume inputs once and hold catalysts}}}
\put(41,40){\vector(0,-1){5}}
\put(4,24){\color{supplyorange!12}\rule{75mm}{11mm}}
\put(4,24){\color{supplyorange}\framebox(75,11){}}
\put(4,24){\makebox(75,11){\parbox{71mm}{\centering\small
Terminal settlement\\Commit outcome and release claims}}}
\put(20,24){\color{supplyorange}\vector(0,-1){10}}
\put(18,19){\makebox(0,0)[r]{\small Yield}}
\put(63,24){\vector(0,-1){10}}
\put(65,19){\makebox(0,0)[l]{\small If recovery}}
\put(1,1){\color{supplyorange!18}\rule{38mm}{13mm}}
\put(1,1){\color{supplyorange}\framebox(38,13){}}
\put(1,1){\makebox(38,13){\parbox{34mm}{\centering\small
Delivered output\\Ready inventory}}}
\put(45,1){\color{black!5}\rule{38mm}{13mm}}
\put(45,1){\framebox(38,13){\parbox{34mm}{\centering\small
Recovery policy\\Fresh retry or fallback}}}
\end{picture}
\caption{Factory admission, physical service, and settlement. Orange
stages manage material and the blue stage uses shared resources. Inputs
commit once. Warm-up and cadence gate admission when enabled.
Unsuccessful fixed-batch attempts yield no output. Configured recovery
starts a fresh attempt.}
\label{fig:factory}
\Description{Orange admission reserves inputs and output capacity. A blue physical workflow consumes inputs once and holds catalysts while using shared resources. Orange settlement commits the outcome and releases claims. Accepted yield becomes ready inventory. A separate labeled branch applies retry or fallback when recovery is configured.}
\end{figure}

Figure~\ref{fig:factory} follows an attempt. Admission reserves its
consumable and catalytic inputs and claims output space atomically.
A failed admission leaves those quantities unchanged. Consumable inputs
commit once at the first physical stage. A fixed-batch attempt deposits
its full output vector after final acceptance, or none on another
terminal outcome. A joint-outcome profile can instead specify realized
yield, quality, and catalyst disposition together with stage durations.

For one material class in one store,
\begin{equation}
\begin{split}
 I(t)={}&I(0)+Y(t)+T_{\mathrm{in}}(t)\\
       &{}-C(t)-T_{\mathrm{out}}(t)-X_{\mathrm{disc}}(t),
\end{split}
\label{eq:inventory}
\end{equation}
where the cumulative terms denote production, transfers, consumption,
and discard. Reservations assign existing material. Incoming claims
reserve future space. Resident quantity plus incoming claims must fit
the store. A transfer reserves both source material and destination space
before route service makes the batch usable.

A material-waiting consumer holds its protected logical allocation but
no injection-stage resources. Its reservation commits at the recipe's
physical consumption point. Earlier cancellation releases the reservation. A later failure retains the consumption. Retries create fresh attempts.
Catalysts remain exclusive through an attempt and return, degrade, or are
lost according to its outcome.

On-demand, maintain-minimum, and autonomous production obey the same
input, storage, resource, and in-flight limits. Producers sharing a store
also share its refill requirement. Launch cadence constrains successful
admissions,
\begin{equation}
 s_{f,n+1}\ge s_{f,n}+\iota_f,
 \label{eq:cadence}
\end{equation}
independently of attempt latency and concurrent capacity. Failed admission
does not advance cadence. Warm-up initializes a deployment using declared
resources before attempts can start. Appendix~\ref{app:protocols} records
its timing, deadline, and fixed-batch conventions.

\subsection{Classical Work, Memory, and Data Movement}
\label{sec:classical}

Classical service needs processors, provider state, scratch space, and
transfer buffers. Memory capacity
limits coexistence, while bandwidth limits progress. Local memory traffic
and wire payload are separate quantities.

Worker slots limit concurrent calls. Input credits limit admitted or
retained history. Output credits limit results awaiting release.
A completed decoder can retain output storage while waiting for the
controller, preventing later publication. Increasing workers cannot
remove that storage constraint.

A two-call example connects these rules. Suppose each call needs a worker,
1,024 bytes of scratch, $1\us$ of compute, and a 1\,MB memory transfer.
If the scratch pool holds only 1,024 bytes, only one call can start even
with two free workers. With 2,048 bytes, both can start. On a shared
100\,GB/s domain with equal weights, their transfers each finish at
$20\us$ rather than the isolated $10\us$. Compute finishes at $1\us$,
but both workers and scratch allocations remain occupied until the
compute/memory join. Section~\ref{sec:validation} checks this latter
timeline against the implementation.

Service profiles identify included costs. A complete measured interval is
charged once. Compute-only service can be combined with explicit memory
traffic, while operation counts use a declared processing rate. With
compute time $C$ and memory time $M$, uncontended serial and overlapping
service take $C+M$ and $\max(C,M)$. Under contention, active transfers
share domain bandwidth by declared weights, with progress updated when
users change. Memory completion releases bandwidth. The worker and scratch
remain held until both branches finish.

Resident bytes, scratch bytes, memory traffic, and wire payload are
specified independently. Copy operations additionally declare their
engines, buffers, endpoints, and traffic for reads and writes. Packing a payload,
copying it across a route, and unpacking it can have different costs and
release points. Only declared phases create work. A payload size alone
does not imply a particular sequence of physical copies.

\begin{table}[t]
\centering\small
\caption{Ownership follows the last dependent phase.}
\label{tab:ownership}
\begin{tabularx}{\columnwidth}{@{}>{\raggedright\arraybackslash}p{.28\columnwidth}X@{}}
\toprule
Resource & Held until \\
\midrule
Worker and scratch & The service interval completes, including its compute/memory join. \\
Provider state & Closing and final output application/commit have drained. \\
Copy engine & Its copy completes. \\
Transfer buffers & The last phase using each physical buffer completes. \\
Logical allocation & Protected computation and issued QEC work have drained. \\
\bottomrule
\end{tabularx}
\end{table}

\subsection{Provider Lifecycle and Usable Feedback}
\label{sec:provider}

A decoder returning an answer and a controller applying that answer are
different events. The provider path separates dispatch and input
conversion, computation, publication, output conversion, collection,
feedback, controller application, and commit. Each enabled phase names
its service, resources, and dependencies. An explicit controller service
replaces the corresponding aggregate interval rather than being added
to it a second time. Information advances at the phase selected by the
logical recipe.

A finite streaming provider also carries state between ordered chunks.
Its workspace is acquired on opening. Input admission respects the
source's contiguous order and releases the associated input credits at
the declared admission point. A chunk can consume input and perform real
decoding while publishing no output. That chunk creates no output-copy
or feedback operation. A final call with empty input may still perform
cleanup and publish the remaining results. Closing and final output
application/commit drain before persistent workspace is released.
Service completion can free the worker while output credits remain held
through collection or commit, according to policy (Table~\ref{tab:ownership}).

Buffer policy determines which available capacity can actually be used.
An ordered provider ring can wait for its next designated slot even when
another slot is free. An any-free policy permits different admissions and
must be selected explicitly. Likewise, reserving output capacity before
service and allocating it only when output is ready place backpressure at
different points. Neither policy turns a buffer slot into a decoder
worker or bypasses the consumer's required feedback frontier.

The CUDA-Q software-injector binding separates three limits. Workers
bound active service. A request-ring slot is held from decoder admission
through feedback handoff. A controller feedback entry remains occupied
until final application. Each has its own capacity and release event.
For staged windows, the separate end-to-end admission policy retains
an input credit through application and reserves feedback space before
service. A free worker cannot admit a request lacking a required credit.
Section~\ref{sec:application-ownership} varies worker and feedback
capacity, then compares reservation and publication policies under
variable service.

Retained provider histories preserve measured order and state transitions,
detector-availability boundaries, and output meanings. The engine imposes resource contention and feedback
ordering. Section~\ref{sec:study-scope} identifies the evaluated paths.

\subsection{Admission, Scheduling, and Placement}
\label{sec:execution}

All work shares one admission rule. A nonpreemptive stage $i$ has
predecessor readiness $a_i$, start $s_i$, and completion $c_i$:
\begin{equation}
 a_i=\max\!\left(\rho_i,\max_{h\in\operatorname{pred}(i)}c_h\right),
 \quad s_i\ge a_i,\quad c_i=s_i+\ell_i.
 \label{eq:operation-time}
\end{equation}
Here $\rho_i$ is release time and $\ell_i$ is realized service, including
any explicitly modeled shared-bandwidth progress. Admission checks the
information and material prerequisites and acquires the entire resource
bundle. For each resource $r$, holders $\mathcal A(t)$ satisfy
$\sum_{i\in\mathcal A(t)}u_{ir}\le C_r$.
Waiting for a bundle does not partially acquire it. Longer-lived
allocations retain their own ownership independently.

Scheduling ranks candidates, and placement chooses feasible endpoints
and routes. Both preserve causal prerequisites and resource capacity. The reference operation-FIFO policy orders ready
work by enqueue time, insertion sequence, and stable identity, with equal
priority for logical, decoder, and factory work. It rechecks complete
bundles and lets later admissible work backfill a blocked candidate.

Execution advances the state $X_n$ through pending events $\mathcal H_n$:
\begin{equation}
 e_n=\arg\min_{e\in\mathcal H_n}(t_e,\phi_e,\sigma_e),
 \qquad X_{n+1}=\mathcal T(X_n,e_n).
 \label{eq:event-transition}
\end{equation}
Integer physical time, event phase, and insertion sequence determine
order. Equal-time phases order topology changes, completions, successor
creation, arrivals, admission timers, dispatch, and accounting. Coupled
ownership and inventory updates commit together. Newly created events
at the current time execute before time advances.

\subsection{Profiles, Outcomes, and Reported Metrics}
\label{sec:outcomes}
\label{sec:metrics}

Profiles supply intrinsic service and outcomes. Queueing follows from
execution. Each profile records its operating coordinate and whether its
measurement covers a stage, batch, or complete interval. Factory outcomes
can jointly retain durations, terminal status, yield, quality, residual
error information, and catalyst disposition:
\begin{equation}
 z=(\boldsymbol\ell,u,\boldsymbol y,\boldsymbol q,e,\boldsymbol c),
 \qquad z\sim p_\theta(z\mid x).
 \label{eq:joint}
\end{equation}
A factory selects an outcome at issue. A decoder selects service and
outcome at admission, with conditional fallback sampled when reached.
Seeds and stable job, window, stage, and attempt identities determine
realizations. Reconsidering blocked work preserves its decision identity.

For completed job $j$, runtime and elapsed QEC-stall fraction are
\begin{equation}
 T_j=t_j^{\mathrm{finish}}-t_j^{\mathrm{submit}},\qquad
 f_j^{\mathrm{stall}}=\frac{|\bigcup_k\mathcal W_{j,k}|}{T_j},
 \label{eq:runtime}
\end{equation}
where $\mathcal W_{j,k}$ are its synchronization-wait intervals.
Their union counts simultaneous waits once. Material-wait sums and
service sums retain concurrent work and therefore do not form additive
parts of runtime. For each protected interval, the last logical operation seals syndrome
generation at $t_{\mathrm{seal}}$. Already issued QEC work, including a
partial final window, must settle before that allocation is released at
$t_{\mathrm{release}}\ge t_{\mathrm{seal}}$. Application completion
$t_j^{\mathrm{finish}}$ includes these releases and any remaining declared
job stages. Autonomous factories can continue replenishment after the
job ends, so final simulator drain can occur later still. Host wall time
measures the cost of simulating this timeline. These endpoints must be
kept separate when reporting performance.

Protected interval $v$ with allocation $L_v$ and actual protected rounds
$R_v^{\mathrm{prot}}$ contributes $N_v=L_vR_v^{\mathrm{prot}}$
logical-qubit-rounds. Ten extra rounds on 32 logical qubits add 320 such
rounds without advancing a logical operation. Footprint $a_v$ converts
this to $a_vN_v$ physical-qubit-rounds. We also record physical qubit-time,
$\sum_v a_vL_v\Delta t_v$, where $\Delta t_v$ is the elapsed protected
interval in seconds. This includes delays between issued rounds.
Controller or extraction contention can therefore reduce the round
count while extending qubit-time. Requested work, executed protection,
and elapsed protection remain distinct. When a compatible
per-logical-qubit-round hazard $p_v$ is supplied, the conditional
independent-hazard proxy is
\begin{equation}
 \log S_{\mathrm{proxy}}=\sum_vN_v\log(1-p_v),\qquad
 P_{\mathrm{proxy}}=1-S_{\mathrm{proxy}}.
 \label{eq:survival}
\end{equation}
Its code, distance, observable, and profile provenance must match. Unsupported hazards remain unavailable. Factory-output quality is
reported separately. Exposure itself requires no error-composition
assumption and directly records the extra protection caused by waiting.

\section{Implementation and Validation}
\label{sec:implementation}

Input preparation records the graph, recipes, topology, profiles, and seeds
in a reproducible request. The engine applies Section~\ref{sec:design}'s
rules and records progress, frontiers, material transactions, and ownership
for analysis.

\subsection{Adapters: From Source Evidence to Execution}
\label{sec:adapters}

Adapters preserve dependencies that gate counts cannot express: independent
consumers, bursts, and feedback paths. Compilation supplies gates and
material demand. The imported graph retains wire order, measurements,
resets, and supported control structure. Logical recipes then bind each
operation to information, states, duration, and resources under the
execution rules of Section~\ref{sec:demand}.

Consider a T gate produced by decomposing a source Toffoli operation.
Compilation supplies its identity and predecessors. Its injection recipe
specifies the required information, one compatible magic state, and the
quantum and controller resources. When the gate is reached, execution
captures its information target and waits for that target, its state, and
its resource claims. Protection continues throughout the wait. The
adapter supplies the causal demand, and the simulator derives its timing
under the chosen architecture.

Artifact bundles for OpenQASM~2/3 and CUDA-Q attach source and compiled-body
identities to the typed execution body. They record compiler versions and
options, hardware and calibration references, and a capability report.
The report separates
preserved information, derived dependencies, supplied assumptions, and
unsupported or missing features. Changing source or compilation requires rebinding.
Serialization checks reload the bundle, re-import its source, and compare
the body and report.

The CUDA-Q Logical bridge exposes two useful entry points. Its logical
stage (P0) supplies operations and value dependencies for binding to
logical recipes. Its QEC realization (P2) and mapped physical graph (P3)
supply lowered work, carrier identities, and dependencies for a physical
binding. The bridge reads typed exports from a pinned compiler build and
checks the source-to-operation mapping. This lets a study retain the
compiler's chosen realization while exploring the shared resources that
serve it. The simulator computes execution times from the supplied
service and ownership rules.

The Stim binding preserves the measurement records needed by each detector.
Provider bindings retain IonQ native gates and Bloqade Lanes movement,
as well as reached Guppy and Selene histories. Recorded outcomes stay paired with the
control decisions they produced, linking adaptive measurements to their
consumers. These physical inputs use the operation identities and timing
contract of Section~\ref{sec:demand}, then share the logical path's admission,
ownership, and event engine. Gate counts and compiler wall time do not
supply physical execution time.

For the complete GF($2^{10}$) multiplier, the adapter reads the quantum
compiler benchmark collection~\cite{quantumcompilerbench} and retains
an exact Clifford+T decomposition of its 100 CCX gates at optimization
level zero. The FTCircuitBench Ising input uses its Clifford+T conversion,
with tolerance $10^{-8}$ for each synthesized rotation
\cite{ftcircuitbench2026}. Unsynthesized rotations cannot silently become
T gates. Source coverage, logical width, compiled-body gates, and the
added preparation/readout wrapper are checked separately. Every T or
T-dagger consumes one state under the selected recipe.
Appendix~\ref{app:application-imports} gives the compilation and gate
accounting for both complete circuits.

\subsection{Graph Preservation and Execution Rules}
\label{sec:validation}

Validation separately checks imported dependencies, execution against
analytic and ownership rules, and component profiles against measurements.
These checks establish the computation, resource composition, and timing
and outcome evidence, respectively.

\paragraph{The input interface preserves the application.}
OpenQASM~2, OpenQASM~3, and CUDA-Q imports of a complete Vedral--Barenco--Ekert (VBE) adder all
produce the same 190-operation body. A common initialization and readout
wrapper gives 210 segments, 266 dependency edges, and graph depth 99.
Gate identities, targets, and graph structure agree across the three
frontends. These checks establish that changing the interface preserves
the application. Compilation tolerance and physical timing remain
separate inputs.

\paragraph{Classical service respects ownership through completion.}
The two-call example in Section~\ref{sec:classical} uses equal weights on
one 100\,GB/s memory domain. The expected completion is
$2\times1\,\mathrm{MB}/100\,\mathrm{GB/s}=20\us$ for each call.
The native resource check completes both at $20\us$ and an independent
1\,MB transfer on a second domain at $10\us$. At $1\us$, computation
has finished but both calls still hold their workers and all 2,048
scratch bytes. All allocations are released after the join.
This checks shared service and release against an analytic timeline
using specified traffic and bandwidth.

A complete VBE copy-path check executes 816 windows. Input copies move
48,960 bytes and charge 97,920 bytes of local read/write traffic.
Output copies move 1,632 bytes and charge 3,264 traffic bytes. These
exact totals check the distinction among wire payload, physical copies,
and memory work. A finite provider check replays 26 ordered calls,
including six initial calls with no output and 20 publishing calls.
Two open/close lifetimes drain completely, including final cleanup.
An adaptive physical-input check executes 23 controller applications,
and every reached correction waits for commit. The source workload and
outcomes are preserved in these checks.

\paragraph{Quantum ownership preserves causal and spatial constraints.}
A fixed-layout VBE execution contains 192 spatial operations and 140
routes over 44 used tiles. All exclusive intervals are disjoint.
Among source-independent gates, 22 pairs wait for a shared route and
start when its holder releases it. In one pair, a T gate holds the route
for $3.3\us$ while a T-dagger waits. Four protection operations execute
during that interval. The trace also records 116 immediate tile reuses
after release. All 63,746 protection rounds and 10,625 decoder windows
settle at termination. This checks resource ownership for the supplied
fixed placement and service assumptions.

Further cases check the fixed target in Figure~\ref{fig:frontier},
exclusive catalyst use, unconstrained fixed-yield production, atomic
material admission, cancellation, and finite-tail shutdown.
Every completed evaluation run must reconcile consumed states,
resolved synchronization requirements, settled QEC work, and released
resources with its request. All 600 application executions pass these
independent checks. Appendix~\ref{app:repro} records the supporting cases
and their evidence files.

\subsection{Component Timing and Outcome Evidence}
\label{sec:grounding}

\begin{table*}[t]
\caption{Component evidence and its role. Coverage is a set of sampled
conditions. Collection hardware and simulated resource capacity are
separate attributes.}
\label{tab:grounding}
\centering\small
\begin{tabularx}{\textwidth}{@{}p{23mm}p{75mm}X@{}}
\toprule
Evidence & Platform and coverage & Use and interpretation \\
\midrule
IBM probes & Fez/Marrakesh, 119 configurations, 140,288 shots.
State and T-state measurements, parity, dynamic operations, postselection,
delays, and repeat anchors. & Physical outcomes and backend/epoch
variation. The logical-memory hazard comes from QEC memory fits. \\
Magic protocols & CUDA-Q on A100 40\,GB, 30 fixed-condition pairs,
20.59 million attempts. Raw T, T teleportation, 15-to-1, CCZ, and
catalytic CCZ-to-two-T. Ideal and independent input Z faults
$10^{-4}$--$10^{-2}$, other operations ideal.
& Joint acceptance, yield, output error, and catalyst outcomes.
RQ studies use raw T and 15-to-1 at input error $10^{-3}$. \\
QEC service & A100 40/80\,GB host platforms, 223 service and 36
sliding-window configurations. CPU PyMatching and GPU BP/tensor-network
paths, rotated surface $d=3,5,7,9,11,13$, sampled physical errors
$3\times10^{-4},10^{-3},3\times10^{-3}$.
& First-use/reused-shape service. Tensor-network evidence at $d=3,5$,
window evidence at $d=5,9,13$. Reference studies use CPU PyMatching. \\
QEC memory & Stim and CPU PyMatching, A100 40\,GB host platforms,
89 configurations, $d=3,5,7,9,11,13$, physical error $10^{-4}$--$10^{-2}$,
separate X/Z observables. & 18 observable-specific parity fits.
The reference uses the $d=3$ Z-memory hazard under independent flips. \\
Multi-family service & Separate retained batch-one coordinate means
for five code families at construction sizes 3, 5, 7. Surface/toric
MWPM and color/HGP BP-OSD matrix service.
& Serial per-block service envelope for the family sensitivity.
Its code-dependent timing mapping is an architectural assumption. \\
\bottomrule
\end{tabularx}
\end{table*}

Table~\ref{tab:grounding} identifies the component evidence used to
parameterize the model. The application reference uses CPU PyMatching
MWPM on rotated-surface memory circuits at distance three, six rounds,
and physical error $10^{-3}$. Four X/Z measurements give a batch-one
mean of $4.702\us$. The controlled reference uses a batch-eight
Z-memory call with mean $23.654\us$. The family studies use separate batch-one coordinates for MWPM or
belief propagation with ordered-statistics decoding (BP-OSD). These bindings identify the evidence behind the
headline comparisons. Window scaling and logical recipe durations are
specified architectural inputs.

PyMatching memory-service lookup models have median absolute percentage
errors of 7.75\% (X) and 7.74\% (Z) on held-out job/condition medians.
Broader reused-service models have errors of 75.43\% for GPU belief propagation and
97.36\% for PyMatching. These are diagnostics on the set used to select
models. Their variation motivates coordinate-specific service bindings
and explicit timing sensitivities. They do not measure accuracy of a
complete quantum--classical machine.

Thirty fixed-condition magic-protocol pairs use separate training and
held-out jobs. The maximum acceptance discrepancy is 0.43 percentage
points, and joint-outcome total-variation distance is at most 0.0124.
These comparisons describe adaptively stopped samples under independent
input faults. An exact-circuit IBM job pair gives held-out log loss
0.306 for a joint outcome law and 0.336 for factorized marginals,
supporting dependence in that physical proxy. Factory protocol schedules
supply physical service durations independently of simulation host time.

For logical-memory evidence, terminal odd parity after $r$ rounds under
independent per-round flips of probability $p$ is
\begin{equation}
 q_r=\frac{1-(1-2p)^r}{2}.
 \label{eq:parity}
\end{equation}
Fits use this observation model and retain the X/Z distinction and
finite-sample uncertainty. Because terminal parity can hide an even
number of flips, converting it into a survival hazard requires that
assumption. Protected exposure is measured directly from the simulation.
Appendix~\ref{app:grounding-accuracy} gives all timing diagnostics and
outcome comparisons, and Appendix~\ref{app:platforms} identifies the
collection environments.

\section{Evaluation Methodology}
\label{sec:application-method}
\label{sec:method}

Controlled demand separates grouping from temporal concentration. Complete
compiled applications test their interaction under algorithmic dependencies.
The studies share the following settings and matched interventions.

\subsection{Workloads and Experimental Roles}

\begin{table*}[t]
\centering\small
\caption{Workloads and their roles. Gate counts describe the compiled
body. Initialization, preparation, and final readout wrappers are
additional. T counts are per modeled job.}
\label{tab:workloads}
\begin{tabularx}{\textwidth}{@{}p{.16\textwidth}p{.15\textwidth}rXrr@{}}
\toprule
Workload & Source & Qubits & Main parameters & Body gates & T states \\
\midrule
Controlled generic & \sys{} generator & 32 & Five sequential circuits,
one shot each. 4,096 Clifford rounds and 128 T consumptions per circuit. Uniform, burst, clustered, and clustered-burst demand. & --- & 640 \\
GF($2^{10}$) multiplier & Quantum compiler benchmarks~\cite{quantumcompilerbench}
& 30 & Complete arithmetic circuit. Exact decomposition of 100 CCX gates. Optimization level zero. One circuit and shot. & 1,747 & 700 \\
1D Ising evolution & FTCircuitBench~\cite{ftcircuitbench2026} & 9 & Complete
five-Trotter-step circuit. Synthesis tolerance $10^{-8}$ per rotation. Alternating basis preparation and Z readout. One circuit and shot.
& 18,045 & 7,110 \\
\bottomrule
\end{tabularx}
\end{table*}

Across five sequential circuits, the controlled job requests 20,480
Clifford rounds ($22.528\,\mathrm{ms}$ at $1.1\us$) and 640 T states.
Initialization and readout bring its synchronization count to 650
(Table~\ref{tab:workloads}).

Uniform and clustered demand use individual consumers. Burst and
clustered-burst use groups of eight independent consumers. Clustering
uses $f=1/2$ and $c=4$, placing half the demand in approximately one
fifth of the Clifford work. Each comparison reuses the same saved graph
and circuit-specific offsets. Six outer CPU, GPU, and controller stages
have specified $1\us$ service each and execute once per job.

The multiplier tests arithmetic with width and T demand comparable to the
controlled job. Ising's much greater T demand and smaller footprint expose
a different balance between supply and protection (Table~\ref{tab:workloads}). Both use
complete compiled circuits with explicit preparation/readout wrappers.
Ising's per-rotation tolerance does not bound whole-application error.

\subsection{Reference Machine and Matched Interventions}

\begin{table*}[t]
\centering\small
\caption{Reference settings shared by the studies and their decoder-mapping differences.
M: component measurement. D: derived quantity. A: architectural assumption or transfer.}
\label{tab:application-machine}
\label{tab:eval-config}
\label{tab:rq1-controls}
\begin{tabularx}{\textwidth}{@{}p{.22\textwidth}Xp{.25\textwidth}@{}}
\toprule
Component & Reference & Intervention or difference \\
\midrule
Protection & Rotated-surface $d=3$, $1.1\us$ round (A). Logical blocks:
32 controlled, 30 multiplier, 9 Ising (D). & Family and size mappings below. \\
Controlled decoder & PyMatching MWPM Z memory, error $10^{-3}$,
batch eight, $23.654\us$ mean (M). & Four sequential calls per window (A). \\
Application decoder & PyMatching MWPM X/Z memory, error $10^{-3}$,
batch one, $4.701786\us$ mean (M). & One serial call per block (A). \\
Windows & Six-round history, zero overlap, 24 outstanding windows (A).
& History, overlap, and limit varied separately. \\
Classical capacity & 32 CPU service slots, one GPU slot,
eight controller slots, operation FIFO (A). & CPU slots 8, 16, 24, 32.
Decoder speed $0.5,1,2,4\times$. \\
Factories & One raw-T/15-to-1 chain, one attempt in flight,
acceptance 0.9844 at input error $10^{-3}$ (M/A). & One, two, or four chains. \\
Inventory & Initially empty. Raw store 64, target 32.
Final store capacity/target 16 (A). & Final capacity 4, 8, 16, 32. \\
Communication & Syndrome readout 150\,ns, links $1\us$ and 50\,GB/s,
controller application 600\,ns (A). & Controller 100, 600, 1,000\,ns. \\
Quantum recipes & Application Clifford steps two rounds.
Initialization, injection, and readout three rounds (A).
& Controlled Clifford intervals follow the generated graph. \\
\bottomrule
\end{tabularx}
\end{table*}

Table~\ref{tab:application-machine} gives the reference settings and the
separate decoder mappings. The controlled study uses four sequential
batch-eight calls per 32-block window. Each application window uses one
serial batch-one call per protected block. CPU slots serve decoding,
frame processing, factory control, and other CPU stages concurrently.
A slot is a unit of modeled service capacity. It is distinct from a host
core, a physical device, and the per-job outstanding-window limit.

One production chain contains raw-T preparation and 15-to-1 distillation.
The raw producer takes $4.95\us$ per state and feeds a 64-state store
with target 32. One distillation attempt can be in flight per chain.
It consumes fifteen inputs and executes QPU/CPU/QPU/QPU stages of
$3.3/0.6/3.3/3.3\us$. The joint outcome profile at input error $10^{-3}$
accepts with probability 0.9844 and yields one state. Replication adds
complete quantum production chains sharing classical capacity and final
inventory. Stores start empty and persist across sequential circuits.
Stage durations, raw cadence, and resource sizes are architectural
assumptions. Acceptance and yield use the component evidence in
Section~\ref{sec:grounding}.

Interventions use the coordinates in Table~\ref{tab:application-machine},
holding dependencies, physical cadence, links, and operation FIFO fixed.
Decoder speed divides service duration without changing concurrency.
RQ2 adds an assumed GPU predecoder with conditional CPU fallback.
Scheduling and telemetry add no modeled service, and the controlled reference
has no barrier deadlines.

\paragraph{Window and service-law studies.}
For reference history $W_0=6$ and measured mean call service $m(W_0)$,
the window sensitivity uses
\begin{equation}
 m(W)=m(W_0)\left(0.2+0.8\frac{W}{W_0}\right).
 \label{eq:window-cost}
\end{equation}
The fixed and history-dependent fractions are assumptions. With overlap
$O$, a full window resolves $W-O$ new rounds, so $n$ sequential calls cost
$n\,m(W)/(W-O)$ per new round. The controlled reference uses $n=4$.
Formation, queueing, transfer, and feedback introduce separate delays.
The service-law studies preserve the expected mean while changing its
distribution. These experiments measure timing and exposure sensitivities
under the specified laws. They do not estimate decoder accuracy on
noisy overlapping histories or hardware tail distributions.

\paragraph{Family and size mappings.}
Both studies examine rotated and unrotated surface codes, toric and color
codes, and hypergraph-product (HGP) codes at construction coordinates
$s=3,5,7$. Surface, toric,
and color use distances $d=s$. HGP uses distances 6, 10, and 14 and
72, 200, or 392 data qubits per logical qubit. Toric blocks encode two
logical qubits. Geometry determines the blocks and syndrome payload
needed for each application's width. Surface and toric service uses
MWPM, and color/HGP uses BP-OSD matrix-service profiles.

These mappings use 256 CPU slots, 256 outstanding windows, serial
batch-one calls per block, and default history $W=2s$. They transfer
their measured coordinate means into the window-service mapping.
Initialization, injection, and readout take $d$ rounds, each $1.1\us$.
Application quantum recipe steps likewise use their mapping's distance-based
service. Factories retain the reference technology. Size therefore changes
block geometry, service, and recipe durations together. Supply and
service-law comparisons are paired within each family and size.
They do not match logical error rates across families.

\subsection{Study Scope and Reporting}
\label{sec:study-scope}

\paragraph{Coupled and fixed protection work.}
We compare the coupled model with an approximation that prescribes
protection work from the nominal operation graph. The approximation
propagates recipe durations through that graph without resource waits,
then rounds each protection prefix up to a complete decoding window.
These prefixes determine information targets and the total protection
budget before simulation. Both models retain finite factories, decoding,
and resource contention. Only the coupled model generates additional
protection while operations wait. Each pair uses the same workload,
machine settings, service laws, and seed. We test 16 and 24 CPU slots
with one and four factory chains on all three workloads.

\paragraph{Worker and feedback ownership.}
The ownership study separates decoder workers from 32 coordination
CPU slots. Its capacity sweep uses 16, 24, and 32 workers with
8, 16, 24, and 48 feedback entries. End-to-end admission reserves
feedback space and retains an input credit through controller
application. The study includes
explicit dispatch, packing, copying, publication, and collection phases.
Dispatch, packing, unpacking, publication, and collection take
50, 20, 10, 100, and 50\,ns, respectively, as architectural assumptions.
A single copy engine moves each input and result through a shared
100\,GB/s memory domain, charging both reads and writes and 1\,ns of
copy-engine service. Decoder calls retain their measured complete
service and reserve 1,024 bytes of scratch. Controller application
remains 600\,ns and is charged once. Input copies carry 180, 54, and
768 bytes for the multiplier, Ising, and controlled workload, respectively.
Their result copies carry four, two, and four bytes. The syndrome buffer
holds 256 rounds, with at most 24 outstanding windows. All cases use
four factory chains. The two resource limits change independently.

The ownership/variability comparison fixes sixteen feedback entries
and uses 16 or 24 workers. Alongside end-to-end credits, reserved
feedback keeps the output reservation but releases consumed input
storage earlier. Ordered publication instead allocates feedback space
when results publish in window order. All three release feedback
entries after application. Each policy is paired with constant service
and the equal-mean tail law of RQ3.

\paragraph{Controller capacity.}
With 24 CPU slots, we vary the shared controller from one to four
and eight slots at one and four factory chains. Its operations include
protection issuance and feedback application. Issuance requires an
available slot even with zero service duration. Controller application
retains its 600\,ns duration. This comparison uses the aggregate
reference path, holding decoder service and quantum recipes fixed.

All performance studies enable
causal logical demand, continuous protection, fixed information targets,
windowed decoding, finite factories and stores, and shared processor,
controller, and link resources. They use aggregate logical recipes.
The ownership study additionally exercises explicit classical phases,
memory traffic, copies, and feedback admission and release.
Persistent provider histories and fixed spatial layouts are checked by
the semantic validation in Section~\ref{sec:validation}.

The controlled reference covers all 200 design coordinates with seeds
101, 202, and 303. Section~\ref{sec:evaluation} reports seed-101 timing
and exact wait traces, plus four additional two-chain speed points.
The three-seed matrix checks the completion and accounting contrasts.
A separate controlled family design
has 56 completions among 60 coordinates spanning five families, three
sizes, uniform/burst demand, and one/two chains. Its window and variability
extensions use the same mappings and seed. Incomplete outcomes remain
marked in the supporting results.

The application provisioning, window, variability, and family studies
contain 100 configuration classes per application, each with seeds
101, 202, and 303. The coupled-model and ownership comparisons use the
same three seeds, as do the ownership/variability and controller studies.
We report the median
and observed range across these three realizations.
Ratios are computed within each seed before
aggregation. Shared controls count once, and ranges are descriptive,
not confidence intervals. Barrier quantiles in the single-seed controlled
study instead describe dependent waits within one job.

Section~\ref{sec:metrics} defines metrics, application completion, and final drain.
Concurrent service and material-wait sums are not runtime partitions.
Unfinished attempts supply no completion estimate. Each study retains
its own controls and service mapping.

Protected exposure is the primary protection-cost measure. Conditional
memory-risk estimates apply Equation~\ref{eq:survival} using a Z-memory
hazard fitted at physical error $10^{-4}$, whereas timing uses $10^{-3}$.
That transfer and the wide hazard interval make absolute risk a sensitivity
result. Appendix~\ref{app:risk-sensitivity} retains the full values.

Controlled wait traces were collected on an Apple M5 Max CPU, and the
three-seed campaigns ran on Perlmutter. Component measurements also use
A100 GPUs and IBM Fez/Marrakesh~\cite{nersc_perlmutter,ibm_qpu_systems}.
Appendices~\ref{app:platforms}, \ref{app:application-imports}, and
\ref{app:application-sweeps} give environments, compilation, and supporting
sweeps. The \texttt{FTWeave-artifact-v2} entry point and reproduction
commands are in Appendix~\ref{app:artifact-entry}.

\section{Controlled Workload Evaluation}
\label{sec:evaluation}

With requested work and T consumption fixed, the controlled studies isolate
three provisioning decisions: how to combine decoder and factory capacity,
how windows and admission affect usable capacity, and when service
variability changes completion. Section~\ref{sec:application-results}
then tests how complete circuits change these decisions, using the
method in Section~\ref{sec:application-method}.

\begin{figure*}[t]
\centering
\includegraphics[width=\textwidth]{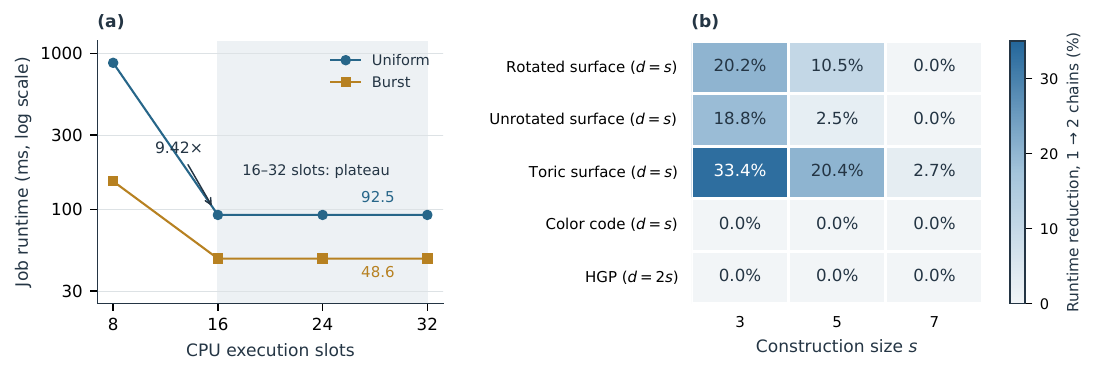}
\caption{RQ1 architectural decisions. (a) Reference rotated-surface
$d=3$ study with one chain. Uniform demand crosses a CPU-capacity
threshold, while both regimes plateau over 16--32 slots.
(b) Separate 256-slot code/decoder study.
Cells show the runtime reduction from one to two chains under burst
demand. Construction size $s$ equals distance for the first four
families and half the distance for HGP. Every cell is a matched
completed pair under the service mapping in Section~\ref{sec:method}.}
\label{fig:rq1-results}
\Description{Left: a logarithmic runtime axis shows the uniform
capacity threshold between eight and sixteen CPU slots and plateaus
thereafter. Right: a five-family by three-size heatmap shows how the
benefit of a second factory changes across code and decoder configurations.}
\end{figure*}

\subsection{RQ1: Joint Architectural Provisioning}
\label{sec:rq1}

\paragraph{Find the classical-capacity threshold.}
Increasing CPU execution slots from eight to sixteen reduces uniform
runtime from $871.1$ to $92.5\,\mathrm{ms}$, a $9.42\times$ speedup
with one factory chain (Figure~\ref{fig:rq1-results}a).
Peak backlog falls from 700 to 145 rounds and p99 synchronization wait
from $1{,}353.6\us$ to $140.0\us$.
Neither 24 nor 32 slots improves completion. More capacity cannot
remove the remaining service and synchronization delay.

\begin{figure*}[t]
\centering
\includegraphics[width=.98\textwidth]{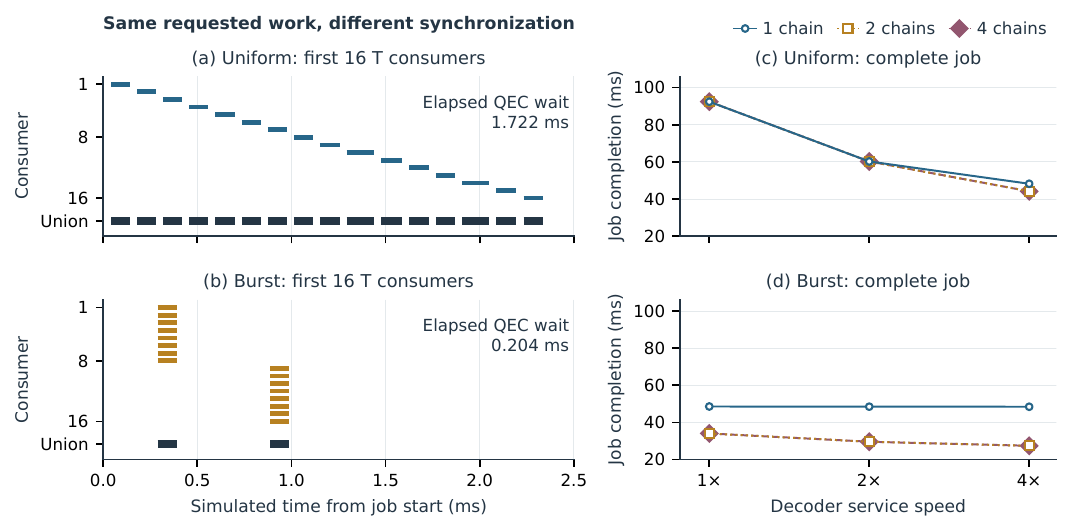}
\caption{Demand grouping changes waiting and the useful upgrade.
(a,b) Saved QEC-wait intervals for the first sixteen T consumers,
covering 512 requested Clifford rounds in each regime. Black strips
count overlapping waits once. Both excerpts use one chain and the same
time scale. (c,d) Complete five-circuit jobs, each consuming 640 T states,
at 32 CPU slots. Only decoder service is scaled, with physical cadence
fixed at $1.1\us$. Two- and four-chain curves coincide.}
\label{fig:rq1-scaling}
\label{fig:rq4-results}
\Description{Uniform consumers encounter sixteen separate QEC waits,
totaling 1.722 milliseconds. Burst consumers share two intervals,
totaling 0.204 milliseconds. Full-job curves show that uniform demand
benefits from faster decoding, while burst demand first needs a second
factory chain to benefit substantially. Four chains match two.}
\end{figure*}

\paragraph{Grouping changes the waiting that must be removed.}
With one chain, burst demand completes 47.5\% sooner than uniform
demand with identical requested work and T consumption.
In Figure~\ref{fig:rq1-scaling}a,b, sixteen consumers encounter separate
QEC waits under uniform demand, but share two intervals under burst
demand. Both excerpts cover 512 requested Clifford rounds.
Counting simultaneous waits once gives $1.72$ versus $0.204\,\mathrm{ms}$
of elapsed QEC waiting.

Across complete jobs, p95 waits are similar: $123.6\us$ for uniform
and $118.7\us$ for burst. Elapsed QEC waits instead differ sharply,
at $67.61$ and $8.99\,\mathrm{ms}$. Shorter protected lifetime reduces
exposure from 2.675 million logical-qubit-rounds under uniform demand
to 1.398 million under burst demand.
Requested operation counts miss this waiting-dependent work.

\paragraph{Compare upgrades to the remaining dependency.}
For reference burst demand, fourfold faster decoding improves completion
only 0.29\%, whereas a second factory chain reduces runtime 29.9\%.
The trace explains why. Faster decoding reduces elapsed QEC waiting
from $8.99$ to $2.61\,\mathrm{ms}$, but leaves the state-supply limit.
The second chain instead eliminates recorded material waiting and
shortens completion. Four chains give the same completion, making two
sufficient for this configuration.

The baseline records $28.44\,\mathrm{ms}$ of summed material waiting.
That sum overlaps the QEC-wait interval, so comparing their magnitudes
alone cannot identify the useful upgrade. Matched interventions establish
which prerequisite delays completion. Clustered bursts retain a small
benefit from four chains, while uniform and clustered singletons remain
unchanged despite production by every chain.

Storage can cover demand peaks. For clustered bursts with one chain, increasing
the output store from four to eight states reduces runtime 8.9\%,
with a smaller additional benefit at sixteen. Four-chain burst workloads
complete equally fast at all tested store capacities.

\paragraph{Re-evaluate supply when the code/decoder changes.}
Figure~\ref{fig:rq1-results}b compares five families at three sizes.
For rotated surface, a second chain reduces burst runtime 20.2\% at
$d=3$, 10.5\% at $d=5$, and zero at $d=7$.
The one-chain $d=7$ case already has zero recorded distilled-state
waiting. Toric configurations show a similar decline, from 33.4\% to
2.7\% benefit over $d=3$ to seven. Color and HGP burst configurations
gain no completion benefit under their selected service envelopes.
All thirteen completed uniform pairs are unchanged.
Useful production capacity thus depends on how quickly protected
computation reaches consumption, as well as on total demand.

\paragraph{Connect provisioning to protection cost.}
\label{sec:rq4}
Figure~\ref{fig:rq1-scaling}c,d varies decoder CPU service by $1/s$ at
fixed physical cadence, capacity, and protocol durations.
Uniform demand benefits mainly from faster decoding. Burst demand first
needs a second factory chain to expose that benefit.
Fourfold decoder speed and four chains give $2.09\times$ uniform
and $1.77\times$ burst speedup. Two chains match four at every plotted speed.

Shorter execution also reduces the protection required along the way.
Data-QPU protection cost falls from 50.01 to 23.94 physical-qubit-seconds
for uniform demand and from 26.14 to 14.81 for burst demand.
Added factory hardware remains a separate resource cost.
With one chain, faster decoding leaves the material limitation and
slightly raises exposure despite the small runtime improvement.

\subsection{RQ2: Windows and Heterogeneous Decoder Capacity}
\label{sec:rq2}

\begin{figure*}[t]
\centering
\includegraphics[width=\textwidth,trim=0 12bp 0 3bp,clip]{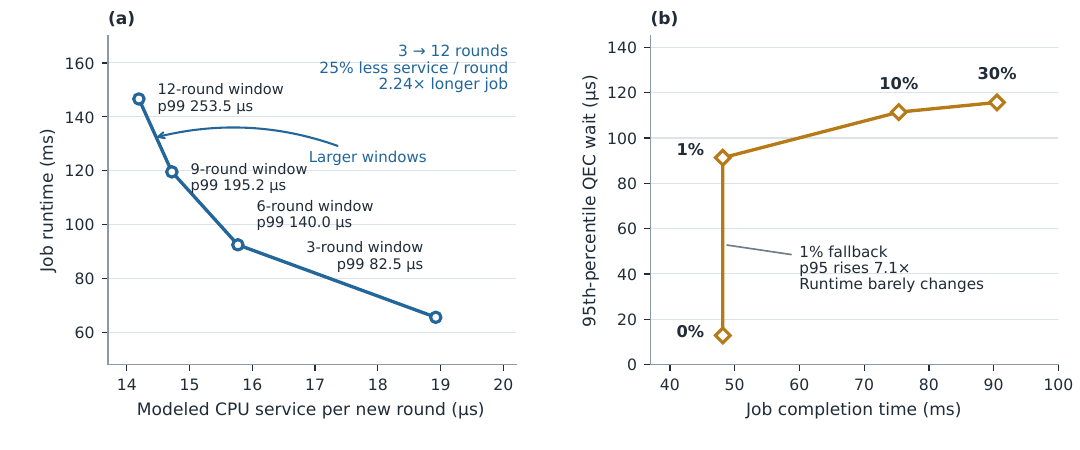}
\caption{RQ2 efficiency and readiness. Left: window size changes service
per new round, uniform runtime, and p99 waiting. Right: GPU predecoding
with CPU fallback changes p95 waiting before substantially changing
runtime. Both panels use one chain, 32 CPU slots, and 24 outstanding windows.
Window scaling, predecoder service, and fallback probabilities are
architectural sensitivities.}
\label{fig:rq2-results}
\Description{Window efficiency can improve while application runtime
worsens. A second panel shows the synchronization-tail effect of
increasing CPU fallback after GPU predecoding.}
\end{figure*}

After service capacity is provisioned, history formation and admission
still determine when information becomes usable.

\paragraph{Less service per round can delay completion.}
With 24 outstanding windows, increasing history length $W$ from three
to twelve rounds reduces assumed service per new round by 25\%, yet
increases uniform runtime $2.24\times$
(Figure~\ref{fig:rq2-results}a).
The cost law in Equation~\ref{eq:window-cost} amortizes service over
longer histories, but those histories take longer to form before the
required information can be decoded. Protected exposure grows from
1.897 to 4.238 million logical-qubit-rounds.

The tradeoff also appears under the family mappings. For color-code
$d=3$ burst demand, six to twelve rounds reduces BP-OSD service per
new round by 10\%, yet raises runtime from 101.5 to 163.0\,ms.
At $d=5$, ten to twenty rounds raises runtime from 97.3 to 154.2\,ms.
For HGP with 72 data qubits ($d=6$), six to twelve rounds raises burst
runtime from 62.6 to 91.2\,ms, and uniform runtime $1.72\times$.
Service efficiency must therefore be assessed together with the
information deadlines imposed by the workload.

\paragraph{Outstanding windows and CPU slots are separate limits.}
With three-round history, reducing the outstanding-window limit from
24 to eight raises uniform runtime from 65.5 to 880.3\,ms at unchanged
CPU capacity. The outstanding limit constrains admitted work, whereas
CPU slots constrain concurrent service. Sizing one without the other
can leave useful execution capacity inaccessible.

\paragraph{Rare fallback affects waiting before completion.}
The GPU predecoder takes $0.05(4m)$ per window, about $4.731\us$. Fallback retains the original CPU path and transport/control
costs. A 1\% fallback rate raises p95 wait $7.1\times$ while completion
barely changes (Figure~\ref{fig:rq2-results}b). At 10\%, runtime is
56.4\% above the zero-fallback case. At 30\%, much of the advantage
over direct decoding disappears. Sixteen and 32 CPU slots give the same
completion in these comparisons. More service capacity cannot remove
the response delay of a request that takes the fallback path.

\begin{figure}[!htbp]
\centering
\includegraphics[width=\columnwidth]{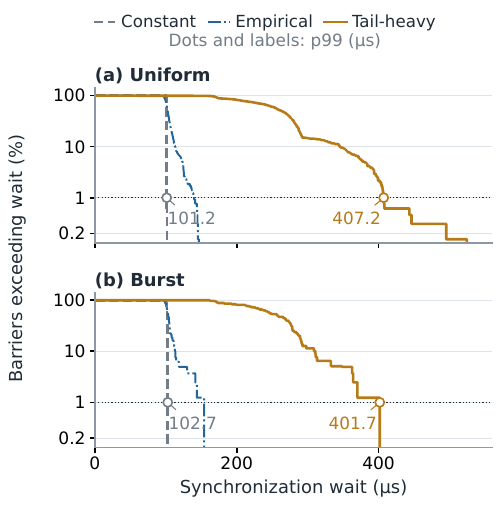}
\caption{RQ3 synchronization-wait exceedance under constant, empirical,
and tail-heavy service. Each curve contains 650 dependent records
within one job. Markers identify nearest-rank p99 and the dashed
line marks 1\%. Labels give p99 in microseconds. Constant and tail-heavy
laws have equal expected mean. Uniform (top) and burst (bottom) expose different runtime
consequences of enlarged individual waits.}
\label{fig:rq3-results}
\Description{Exceedance curves reveal enlarged synchronization tails
under an equal-mean variable profile for both uniform and burst demand.}
\end{figure}

% Declare the next section's opening figure early for two-column placement.
\ProtectionComparisonFigure

\subsection{RQ3: Equal Means, Different Synchronization}
\label{sec:rq3}

RQ3 compares constant service $m$ with $0.5m$ at probability 0.9 and
$5.5m$ at probability 0.1, preserving expected service and execution rules.
Realized uniform CPU-call means differ by only 0.056\%.
Figure~\ref{fig:rq3-results} plots wait exceedance, with p99 at the
1\% crossing.

\paragraph{A delayed frontier generates further work.}
The tail-heavy law increases uniform and clustered runtime
$2.16\times$ and $2.11\times$.
Uniform peak backlog grows from 92 to 479 rounds despite approximately
unchanged mean decoder utilization. A slow outstanding window delays
the contiguous frontier, extending the protected interval and generating
more work. Exposure rises from 2.596 to 5.589 million logical-qubit-rounds.
Mean service and utilization therefore miss a substantial completion cost.

\paragraph{Another dependency can hide that delay at completion.}
Burst runtime rises only 0.86\%, even as p99 grows $3.91\times$ and
elapsed QEC waiting increases from 8.65 to 21.47\,ms.
Material waiting decreases as decoding occupies more of the interval.
State supply masks much of the readiness change in total runtime.
For HGP $d=10$ burst demand, the same equal-mean law applied to BP-OSD
raises runtime from 112.1 to 142.6\,ms, or 27.2\%.
Both HGP runs have zero recorded material waiting, so supply does not
hide variability as in the reference burst case.
These matched contrasts explain why service tails can require attention
even when a completion curve remains flat.

%\FloatBarrier
\section{Complete-Application Results}
\label{sec:application-results}

Complete circuits turn the controlled mechanisms into three architectural
questions. Which upgrade does the application need? Can it use the
provisioned capacity? When do window, service, or code choices change
that recommendation? We first locate the limiting dependency, then test
ownership and feedback, and finally vary the execution mapping.

\subsection{Which Upgrade Does the Application Need?}
\label{sec:application-upgrade}

\subsubsection{Fixed protection work can recommend an ineffective upgrade.}
\label{sec:application-coupling}

Figure~\ref{fig:protection-comparison} compares four choices from the
same multiplier baseline: 16 CPU slots and one factory chain.
The upgrades raise CPU capacity to 24 slots, supply to four chains,
or both. Under fixed protection
work, factories alone give $3.55\times$ speedup and CPU capacity adds
no benefit. Under coupled protection, factories alone barely help,
CPU-only gives $2.45\times$ speedup, and both give $6.06\times$.
The models imply different upgrade sequences.

The CPU upgrade exposes the supply bottleneck, as in RQ1.
In the seed-101, 16-slot, four-chain case, accumulated decoder service
is 2,078\,ms under coupled protection and 21.2\,ms under fixed work.
These sums across calls explain why fixed work misses the classical
bottleneck. They measure service work, separately from elapsed completion.

Agreement also depends on the current limiting path. With one chain,
the Ising approximation differs by only 0.02\%. With four chains,
it underpredicts completion by 30.6\%. The controlled workload differs
by about 1.55\% across these settings.
An approximation can agree in a supply-limited configuration and still
misidentify the benefit of the next upgrade.

\ApplicationProvisioningFigure

\subsubsection{Classical capacity determines the value of more factories.}
\label{sec:application-joint}
\label{sec:application-capacity}
\label{sec:application-matrix}

Having established that continuing protection changes the useful upgrade,
the broader sweeps locate capacity plateaus and test faster service.
The multiplier has 700 T gates and 30 protected logical blocks, whereas
Ising has 7,110 T gates and nine blocks. Each reference window requires
one serial decoder call per block. At the nominal six-round cadence,
30 calls of $4.702\us$ arrive per $6.6\us$ for the multiplier,
requiring an average 21.37 slots of decoder service. Ising requires
6.41 slots. The controlled mapping's four batch calls require 14.34.
These input-derived loads explain the respective 24-, eight-, and
16-slot plateaus. T count determines state demand, while protected
width and service per window determine this classical load.

For the multiplier, 24 slots are the smallest tested capacity reaching
the 24--32-slot plateau, at each factory count
(Figure~\ref{fig:application-provisioning-matrix}c).
Extra factories give less than 0.05\% benefit at eight or sixteen slots.
The traces corroborate this transition: at eight slots,
QEC-wait union is 260.1\,ms with no recorded material waiting.
At 24, QEC waiting falls to 47.1\,ms and material waiting appears.
At that point, further CPU slots are unnecessary and supply becomes
a useful next upgrade.

Ising reaches its capacity plateau at eight slots, the lowest tested.
One to four chains reduces completion from 536.1 to 214.6\,ms at
every tested CPU capacity. Its next useful classical upgrade is faster
service: with four chains, fourfold decoder acceleration reduces
completion a further 36.7\%, to 135.9\,ms.
The speed sweeps separate response latency from service concurrency
(Figure~\ref{fig:application-provisioning-matrix}a,b).
Even at 32 slots, half-speed decoding holds multiplier completion at
176.2\,ms with either one or four chains. At twice the original speed,
these settings separate to 52.8 and 14.8\,ms.
Once capacity saturates, shorter service addresses the remaining response path.

At 32 slots with one chain, state supply limits both applications, and fourfold
faster decoding improves completion by less than 0.5\%.
Four chains together with fourfold decoding instead give about
$3.94\times$ paired speedup for both applications and reduce protected
exposure by approximately 74.7\%. The useful combination removes waiting
and the protection generated during that waiting.

\FloatBarrier
\subsection{When Can Provisioned Capacity Be Used?}
\label{sec:application-usable-capacity}

The capacity thresholds above assume that requests can reach the workers
that serve them. We now separate those workers from feedback storage and
controller service to test which resources keep that capacity usable.

\OwnershipInteractionFigure
\OwnershipTailFigure

\subsubsection{Extra workers need enough feedback storage.}
\label{sec:application-ownership}

With 16 feedback entries, increasing decoder workers from 16 to 24 leaves
multiplier completion at 105.9\,ms (Figure~\ref{fig:ownership-interaction}).
Increasing entries alone to 24 gives 102.4\,ms. Increasing both gives
21.5\,ms, a $4.93\times$ speedup.

Feedback entries remain reserved beyond decoder computation through
controller application. Occupied entries block admission, so extra workers
help only when storage permits more requests. Memory, copying,
coordination, and admission settings stay fixed
(Section~\ref{sec:study-scope}).

The full grid identifies the useful capacity: 24 workers and 24 entries
reach the multiplier plateau. Neither 32 workers nor 48 entries improves
it. The controlled job needs sixteen entries, completing in 89.8\,ms
versus 723.9\,ms at eight. Ising stays at 214.6\,ms throughout the grid.
The useful storage provision therefore depends on the same workload
structure that sets decoder demand.

\paragraph{Service tails expose the retained ownership.}
Figure~\ref{fig:ownership-tails} holds 24 workers and sixteen feedback
entries fixed. Equal-mean tails increase controlled completion
$7.09$--$9.64\times$ across the three policies. The largest penalty,
with reserved feedback, is $9.63$--$9.64\times$ across all three seeds.
Sixteen and 24 workers give identical completion for this policy.

The seed-101 reserved-feedback case explains why extra workers do not
help. Protection work rises $4.91\times$, yet average active
decoder service falls from 14.3 to 7.3 slots. Recorded decoder-stage
queue wait is zero in both cases. Waiting accumulates around admission
and the feedback frontier. More work coexists with idle service
capacity because occupied entries prevent that capacity from being used.

For the multiplier, ordered publication remains fastest under both laws:
21.5\,ms with constant service and 115.1\,ms with tails. Its larger
$5.36\times$ slowdown reflects its faster baseline. Ising's three
policies instead incur similar 51--52\% penalties. Provisioning must
therefore consider ownership under variable service, as well as the
constant-service capacity threshold.

\ControllerCapacityFigure

\subsubsection{Controller capacity becomes useful after supply improves.}
\label{sec:application-controller-capacity}

Once storage permits admission, controller service can limit usable
feedback. At 24 CPU slots,
increasing controller capacity from one to eight slots barely changes
multiplier or Ising completion with one factory chain
(Figure~\ref{fig:controller-capacity}a). With four chains, it reduces
multiplier completion from 25.3 to 21.5\,ms and Ising from 264.6 to
214.6\,ms, improvements of 15.1\% and 18.9\%
(Figure~\ref{fig:controller-capacity}b). Supply hides the controller
benefit until enough states are available.

The upgrade also increases executed protection rounds by 39.0\% and
28.7\%, respectively, while reducing physical qubit-time by 15.2\%
and 18.9\%. Controller contention delays protection issuance as well
as feedback application. Relieving this contention permits more rounds
within a shorter protected lifetime. For seed-101 Ising, summed
protection-issue waiting falls from 97.9\,ms at one slot to zero at eight.
Fewer executed rounds alone would therefore select the slower machine.

\subsection{When Does the Provisioning Choice Change?}
\label{sec:application-choice-sensitivity}

The preceding studies size resources for a given execution mapping.
We next vary history windows, decoder service distributions, and code
families and sizes to test when that provisioning decision changes.

\ApplicationWindowFigure

\subsubsection{Window efficiency and feedback delay compete.}
\label{sec:application-windows}
\label{sec:application-window-work}

The RQ2 tradeoff selects six-round history with zero overlap for the
multiplier, completing in 53.0\,ms versus
86.4\,ms at three rounds and 55.3\,ms at twelve
(Figure~\ref{fig:application-window-work}). Twelve rounds reduce
decoder work by 6.25\% relative to six, but raise QEC waiting from
47.1 to 54.5\,ms. Selecting by service work would choose the slower window.

Overlap changes the two applications differently. With six-round history,
increasing overlap from zero to three rounds raises multiplier decoder
work $3.31\times$ and completion time $2.97\times$.
For Ising, the same change doubles decoder work while completion stays
near 536.1\,ms. Its supply bottleneck hides the extra service at completion.

Multiplier QEC waiting grows from 47.1 to 154.8\,ms. The extended protected
lifetime generates further work, amplifying the direct cost of overlap.
Outstanding capacity is a separate constraint: with three-round history,
reducing its limit from 24 to eight raises completion from 86.4 to 262.6\,ms.

\ApplicationVariabilityFigure
% Queue the final wide panel early so it stays beside the family/size study.
\ApplicationFamilyFigure

\subsubsection{Equal mean service can hide large mapping sensitivities.}
\label{sec:application-variability}
\label{sec:application-tail-mappings}

RQ3 showed that supply can hide decoder variability at completion.
Color and HGP expose the largest application penalties
(Figure~\ref{fig:application-tail-mappings}). At size coordinate five,
the tail-heavy law raises multiplier completion 56.4\% under color
and 52.9\% under HGP. The Ising penalties are 90.1\% and 80.9\%.
The same variability has little effect on reference Ising completion.

Moderate variability selects $0.5m$ or $1.5m$ equally. The tail-heavy law
is the equal-mean law of RQ3. Both preserve expected service $m$.

Reference controls show where capacity or supply absorbs the effect.
The multiplier's tail penalty falls from 5.37\% at 16 CPU slots to
0.76\% at 24.
Ising's median paired slowdown at 16 slots is only 0.04\%, even as
median peak backlog grows from 45 to 154 rounds.
Supply hides the delay while backlog accumulates. Under color and HGP,
information service constrains progress. Their provisioning therefore
requires checking service tails even when the reference mapping is insensitive.

\subsubsection{The supply benefit changes across families and sizes.}
\label{sec:application-family-matrix}

A second factory helps the surface and toric mappings much more than
color and HGP (Figure~\ref{fig:application-family-matrix}).
This extends the controlled family/size comparison in
Figure~\ref{fig:rq1-results}b to both complete applications.
At coordinate five, the multiplier's surface/toric mappings gain
substantial completion reductions, while color is unchanged and HGP
improves by less than 1\%.
Ising follows the same distinction: surface/toric completion falls
from about 536\,ms to 282--325\,ms, while color and HGP remain
unchanged. Large T demand alone does not make another factory useful.

Size also changes the recommendation within a family.
For the unrotated-surface multiplier, the second-chain benefit falls
from 45.0\% at $d=3$ to 13.0\% at $d=7$.
The mapped geometry, decoder service, and recipe durations determine
how quickly the application can consume states. A supply choice made
at one size therefore needs reevaluation as that path changes.
Absolute times appear in Figure~\ref{fig:application-family-absolute}.

%\FloatBarrier

\section{Architectural Implications}
\label{sec:discussion}

Provisioning requires three checks. First, examine completion, protection
work, and elapsed qubit-time together: supply can hide additional work,
while controller relief can shorten lifetime despite more issued rounds.
Second, identify the unmet prerequisite at waiting operations. Summed
material waits can overlap information waits, so their magnitudes alone
do not rank upgrades. Third, change the suspected constraint and measure
completion again.

Capacity, latency, and ownership require different interventions.
Additional workers help calls competing for service. Faster calls help a
serial response even when workers are idle. Storage must permit admission
and remain available through the required feedback phase. The worker/storage
and controller studies show why counting processors alone is insufficient.
Recheck the limiting dependency after each upgrade: changing supply or
feedback can change both the remaining bottleneck and the protection work
created before completion.

The next useful measurements are those that could change a provisioning
decision. For example, operation-specific decoder costs could shift a
window tradeoff, and a different extraction or reset schedule could
change the interval available for feedback. FTPrimitiveBench provides
annotated memory, Hadamard, surgery, and S-gate circuits for developing
such profiles \cite{ftprimitivebench}. Its integration remains future
work. The present results already identify where additional information
would matter: near a capacity threshold or where competing prerequisites
finish at similar times.

\section{Conclusion}
\label{sec:conclusion}

\sys{} connects logical progress, continuous protection, finite
magic-state supply, and classical service on one event timeline. An
operation waits for the information it required when reached, while
later protection rounds continue to create work. Explicit ownership
keeps quantum allocations, workers, memory, copies, and provider state
live until their dependent phases finish. Circuit adapters preserve
compiled operations and dependencies before binding them to a chosen
architecture.

The controlled and complete-application studies show
why provisioning must follow the dependency that limits progress.
Fixing protection work in advance can predict a useful factory
upgrade where coupled execution shows almost no benefit. Factories can
have little value until enough classical capacity is available, and
extra decoder workers can remain ineffective until feedback storage
also grows. Service tails can intensify this ownership constraint.
Controller capacity can shorten protected lifetime while permitting
more protection rounds to execute. Longer windows can reduce decoder work while delaying
completion. Equal-mean service laws can produce substantially different
outcomes when information readiness limits the application. Joint
factory and decoder provisioning can shorten execution and avoid the
protection work that waiting would otherwise require.

By exposing these interactions, the simulator supports concrete choices
about quantum supply, classical capacity, and the timing of usable
feedback. Researchers can compare resource combinations, locate the
remaining waits, and identify which component measurements could change
a design decision before a complete tightly integrated machine exists.

\label{main-text-end}
\bibliographystyle{ACM-Reference-Format}
\bibliography{references}
\clearpage
\setcounter{dbltopnumber}{2}
\appendix
\section{Supporting Validation and Grounding Details}
\label{app:repro}

The supplement has three purposes. Sections~\ref{app:repro},
\ref{app:application-imports}, and~\ref{app:platforms} provide validation,
source accounting, and execution environments. Section~\ref{app:grounding-accuracy}
and Section~\ref{app:risk-sensitivity} explain component accuracy and
conditional reliability. Sections~\ref{app:demand-policy} and
\ref{app:application-sweeps}, together with the intervening controlled
sweeps and model details, retain the complete supporting comparisons.
The main text contains the model rules, settings, and evidence needed
for its architectural conclusions.

\subsection{Artifact Entry Point}
\label{app:artifact-entry}

The companion is named \path{FTWeave-artifact-v2}.
Its root file \path{ARTIFACT.md} gives checksum verification and commands
to rebuild Figures~\ref{fig:protection-comparison}--\ref{fig:application-family-matrix}.
The package contains all 600 application rows, matched-comparison
inputs and summary identities, plotting code, and compact model-validation
records. Data retain exact values where
prose rounds them. The entry file also records access to simulator source,
full execution inputs, and calibration archives, which are separate from
this figure-reproduction package.

\subsection{Supporting Semantic Cases}

Table~\ref{tab:semanticcases} groups checks by semantic boundary.
Completed controlled-family runs also verify request and result identities, the
declared 640 T consumptions and 650 synchronization records, and
participation by every configured production chain.

\begin{table}[b]
\caption{Semantic validation cases complementing the main-text invariants.}
\label{tab:semanticcases}
\centering
\small
\begin{tabularx}{\columnwidth}{@{}>{\raggedright\arraybackslash}p{24mm}>{\raggedright\arraybackslash}X@{}}
\toprule
Boundary & Validation coverage \\
\midrule
Demand and logical execution & Causal predecessors, concurrent T/CCZ segments, joins, allocation bounds, and exact source-layer provenance \\
Protection lifetime & Continued extraction during waits, finite horizon sealing, correct tail drain, and physical time-unit conversion \\
Synchronization & Fixed captured target, decoded/applied/committed selection, exact deadline classification, and overlapping-stall accounting \\
Decoder service & Window/overlap ranges, stage routing, parallel capacity, keyed latency realization, and terminal-window behavior \\
Factory admission & All-or-nothing input reservation and output claims, bounded in-flight work, and no cadence advance on failed admission \\
Material settlement & Unique consumption, realized output bounds, quality compatibility, transport conservation, and exact catalyst disposition \\
Failure and recovery & Before/after-consumption cancellation, timeout ordering, fresh retry identities, compatible fallback, and terminal cleanup \\
Observation and execution & Agreement of committed state and metric projections, including overlapping waits and terminal accounting \\
\bottomrule
\end{tabularx}
\end{table}

\paragraph{Model-check evidence.}
The companion data identifies the native shared-memory test and its
analytic timeline. Checks CP3.1, CP4.2, CP5.1, and CP5.2 provide the
memory, copy, provider, and adaptive-feedback results in
Section~\ref{sec:validation}. The fixed-layout example uses the MP3.1
complete-adder reservation trace. The package includes these compact
records and a source map. Each record preserves the assumptions and
checks for its execution example.

\subsection{Workload and Protocol Conventions}
\label{app:protocols}

The six outer workflow stages are CPU preparation, GPU processing,
CPU transformation, and controller dispatch before the circuits, followed
by CPU receive and postprocessing. Each has assumed $1\us$ service
and executes once per job. Quantum initialization and readout execute
for every circuit-shot pair.

Factory warm-up uses declared resources but consumes no magic inputs
and produces no output. Concurrent requests share that initialization,
and the deployment remains warm through the experiment. Completion
deadlines start at successful admission, separating warm-up and queueing
from attempt latency. Completion wins an equal-time deadline tie.
Fixed-batch protocols verify at the final stage, while early-abort
protocols require an explicit earlier stopping rule. Catalysts remain
exclusively held and return unchanged under the fixed-batch contract.
Joint outcomes can instead partition them into returned, degraded,
and lost quantities, separately from newly produced output.

On-demand, maintain-minimum, and autonomous launch policies share the
same input, storage, in-flight, and resource limits. Producers share a
refill requirement. Assignments respect each producer's available
in-flight capacity. Equal candidates are served round-robin, and
unassigned demand remains available for later admissions.

\section{Application Import and Evidence Accounting}
\label{app:application-imports}

\paragraph{GF($2^{10}$) multiplier.}
The source file is \path{gf2^10_mult.qasm}. We retain commit
\texttt{a88111a9} of the quantum compiler benchmark collection
\cite{quantumcompilerbench}. It contains 30 qubits, 238 H gates, 100 CCX
gates, and nine CX gates. Qiskit 2.5.2, optimization level zero and
transpiler seed zero produce 609 CX, 438 H, 400 T, and 300 T-dagger
gates: 1,747 body operations. No arbitrary-angle synthesis is needed.
Adding 30 initialization and 30 readout operations gives 1,807 demand
segments. The retained graph has 2,386 dependency edges. Its node-depth
is 380, which is a graph property rather than a physical runtime.

\paragraph{Ising evolution.}
The retained FTCircuitBench source is
\texttt{hamiltonians\_5trotter/ising\_1d\_9q.qasm}. It has nine qubits,
90 CX gates and 90 U3 gates. The retained compilation uses Qiskit 2.0.2,
NWQEC 0.1.2 and FTCircuitBench's C++ Clifford+T synthesis path with
per-rotation tolerance $10^{-8}$. The body contains 7,290 H, 7,110 T,
3,240 S, 225 S-dagger, 90 CX and 90 Z gates, totaling 18,045. Nine
initializations, X preparation on qubits 0, 2, 4, 6, and 8, and nine
final Z readouts give 18,068 demand segments. This wrapper is an explicit
experimental input. It does not establish a particular physical
Hamiltonian-observable accuracy target.

\paragraph{Service transfer.}
The application's batch-one decoder coordinate averages four retained
six-round, distance-three X/Z memory measurements. It differs from the
historical generic study's batch-eight Z-memory mapping. The application
request serializes one call per protected block in each window. Its
communication payloads are derived from those blocks. Shared references
are reused exactly across interventions. Raw host execution time and
compiled graph depth are not substituted for modeled completion.

\paragraph{Evidence cohorts.}
The original frozen application set contains 588 unique executions, 294
per application, across 196 configuration classes. A matched follow-up adds
two-chain decoder speeds of $0.5\times$ and $2\times$ for both applications,
with the same three runtime seeds and original native binary. The combined
cohort contains 600 executions and 200 configuration classes, with 300
executions and 100 classes per application. The analysis retains completion, terminal drain, QEC settlement,
resource release, exact request identity, and source/result digests.
The main figures select matched comparisons from the combined set and reuse
reference points. The controlled-demand reference covers its full
200-coordinate grid. The separate family results retain their own
cohort boundaries and incomplete-outcome markers.

\subsection{Hardware and Software Environments}
\label{app:platforms}

The controlled generic cohort runs on the 18-core CPU of an Apple M5 Max MacBook Pro
with 128\,GB unified memory. Its 40-core integrated GPU is not used
for these simulations. Simulated GPU capacity is supplied through
service profiles, independently of host execution resources.

The complete-application cohort executes on Perlmutter CPU hosts.
Its simulator host and measured decoder-service environment are separate
from the modeled CPU-slot count. Import preparation uses Qiskit 2.5.2
for the multiplier and Qiskit 2.0.2 with NWQEC 0.1.2 for Ising.

IBM grounding uses the Fez and Marrakesh backends, each with a
156-qubit Heron r2 processor~\cite{ibm_qpu_systems}. Retained records
cover 100 configurations on Marrakesh and 19 on Fez.

NVIDIA grounding uses one A100-SXM4 GPU per task on NERSC Perlmutter.
Each node has a 64-core AMD EPYC 7763 CPU, 256\,GB DDR4 memory,
and four 40/80\,GB A100 GPUs~\cite{nersc_perlmutter}.
CPU decoder service is measured on the host and GPU service on the
allocated A100. The node's Slingshot 11 interconnect is distinct from
the modeled link latency and bandwidth.

Table~\ref{tab:software-environments} records the collection environments.
The CUDA 13.2 stack supplies the magic-protocol and QEC evidence.
Earlier family-profile measurements use a separate Python 3.12.11,
PyTorch 2.8.0+cu128, and CUDA 12.8 environment on A100 40\,GB nodes.
Those exports identify Stim, PyMatching, qecsim, and LDPC decoder paths,
but do not retain every decoder package's individual version.

Native QEC builds additionally retain source revisions. Tensor-network
software strata remain separate in the accuracy analysis. IBM records
include backend identity, calibration epoch, transpiled circuits, and
provider-job associations. Library availability does not calibrate the
assumed window and predecoder service laws used in RQ2.

\begin{table}[t]
\caption{Retained grounding software. Versions identify collection
and qualification environments, not simulated service times. RQ2's
GPU predecoder remains an assumed sensitivity law.}
\label{tab:software-environments}
\centering\small
\begin{tabularx}{\columnwidth}{@{}>{\raggedright\arraybackslash}p{23mm}>{\raggedright\arraybackslash}X@{}}
\toprule
Evidence path & Retained software versions \\
\midrule
NVIDIA runtime & Python 3.12.14, CUDA Toolkit package 13.2.1, driver 580.159.04 \\
Quantum protocols & CUDA-Q 0.15.1~\cite{cudaq}, cuQuantum Python 26.3.2, cuStateVec 1.13.1, CuPy 13.6.0 \\
QEC and memory & CUDA-Q QEC 0.7.0~\cite{cudaqqec}, Stim 1.16.0~\cite{stim2021}, PyMatching 2.4.0~\cite{higgott2025} \\
Independent cross-checks & cuStabilizer 0.3.0 \\
Tensor-network paths & cuTensorNet 2.12.2~\cite{cuquantum2023}, cuTensor 2.7.0, quimb 1.15.0, cotengra 0.8.2 \\
Realtime qualification & TensorRT 10.16.1.11, PyTorch 2.14.0+cu132, pinned CUDA-Q QEC native builds \\
IBM probes & Qiskit 2.5.2, Qiskit Aer 0.17.2, Qiskit IBM Runtime 0.49.0 \\
Data handling & NumPy 2.2.6, SciPy 1.18.1, h5py 3.16.0, PyArrow 23.0.1 \\
\bottomrule
\end{tabularx}
\end{table}

\subsection{Predictive Diagnostics and Their Domains}
\label{app:grounding-accuracy}

The QEC fitting set contains 348 configurations, comprising
223 decoder-service, 89 memory, and 36 sliding-window configurations.
There are 278 training and 70 held-out configurations. Provider jobs
do not cross these splits. The fitting process produces twenty timing
profiles and eighteen conditional round-parity profiles across nine
fixed-condition groups, keeping X and Z observables separate.

For timing, let $o_h$ and $\widehat o_h$ be the medians of observed
and predicted call durations for held-out provider-job/condition pair
$h$ within one profile. Conditions include distance, batch size, noise,
round count, overlap, and physical cadence. We report
\begin{equation}
 \epsilon_h=100\frac{|\widehat o_h-o_h|}{o_h},
 \qquad \epsilon_{50}=\operatorname{median}_h\epsilon_h .
\end{equation}
Each pair receives equal weight. The 90th percentile linearly
interpolates ordered errors. These statistics measure central-tendency
agreement, with full-distribution accuracy and uncertainty requiring
separate tests. Candidate selection uses the same held-out likelihood
scores, so this is a model-selection set, not an untouched test set.

Table~\ref{tab:timing-accuracy} retains all twenty strata. CPU memory
lookup models have lower median errors than broad service models that
combine distance, noise, batch, and lifecycle conditions. Likelihood-based
selection can still give poor location predictions at individual
coordinates. Architectural comparisons therefore retain their specific
service bindings and separate assumed scaling from measured timing.
First-use, reused-shape, and loading costs remain distinct.

\begin{table*}[t]
\caption{Held-out timing diagnostics for every fitted service stratum.
$H$ counts provider-job/condition pairs. Errors are percentages.
Memory, service, and window identify collection protocols. TN-A and
TN-B retain separate pinned software strata. Lookup, empirical, Gamma,
and lognormal identify the selected statistical models.}
\label{tab:timing-accuracy}
\centering\small
\begin{tabular}{@{}llllrrr@{}}
\toprule
Decoder / basis & Collection & Phase & Model & $H$ & Median error & p90 error \\
\midrule
PyMatching X & Memory & First use & Lookup & 17 & 13.64 & 52.27 \\
PyMatching X & Memory & Reused & Lookup & 12 & 7.75 & 42.65 \\
PyMatching X & Memory & Load & Lookup & 17 & 9.43 & 64.24 \\
GPU BP Z & Service & First use & Lookup & 31 & 51.19 & 97.52 \\
GPU BP Z & Service & Reused & Lookup & 18 & 75.43 & 677.07 \\
GPU BP Z & Service & Load & Empirical & 18 & 51.19 & 492.50 \\
PyMatching Z & Window & First use & Lookup & 4 & 4.26 & 8.26 \\
PyMatching Z & Window & Load & Lookup & 4 & 23.42 & 26.73 \\
PyMatching Z & Memory & First use & Lookup & 17 & 11.44 & 51.52 \\
PyMatching Z & Service & First use & Lookup & 34 & 48.30 & 140.10 \\
PyMatching Z & Memory & Reused & Lookup & 12 & 7.74 & 42.41 \\
PyMatching Z & Service & Reused & Empirical & 22 & 97.36 & 99.92 \\
PyMatching Z & Memory & Load & Lookup & 17 & 27.89 & 62.84 \\
PyMatching Z & Service & Load & Lookup & 22 & 24.13 & 202.62 \\
TN-A Z & Service & First use & Gamma & 6 & 148.50 & 4524.01 \\
TN-A Z & Service & Reused & Empirical & 4 & 76.27 & 89.43 \\
TN-A Z & Service & Load & Lognormal & 4 & 20.41 & 48.52 \\
TN-B Z & Service & First use & Empirical & 6 & 2869.69 & 6622.33 \\
TN-B Z & Service & Reused & Lookup & 3 & 6.44 & 11.56 \\
TN-B Z & Service & Load & Empirical & 3 & 94.02 & 94.10 \\
\bottomrule
\end{tabular}
\end{table*}

\paragraph{Fixed-condition magic outcomes.}
Training and held-out empirical laws use the same complete outcome
tuple, including acceptance, measurement bits, output count, output
error, and catalyst diagnostics. For outcome $z$, their total-variation
distance is
\begin{equation}
 d_{\mathrm{TV}}=\tfrac12\sum_z
 |\widehat p_{\mathrm{train}}(z)-\widehat p_{\mathrm{held}}(z)| .
\end{equation}
Every protocol has six fixed-condition pairs, including the ideal
condition. The thirty pairs span 28 provider jobs, with no job crossing
the training/held-out split. The median TV distance is 0.000916,
with maximum 0.0124. Table~\ref{tab:magic-agreement} also reports acceptance
agreement. Protocols with deterministic acceptance can still have
nonzero output-error disagreement. These are descriptive sample
comparisons under adaptive stopping, with one provider job per side
at each condition. They do not establish between-job uncertainty,
unseen-noise interpolation, or zero rare-event probability. The noise
model injects independent input Z faults and leaves other operations ideal.

\begin{table*}[t]
\caption{Fixed-condition magic-protocol agreement. Counts sum over
six pairs per row. Acceptance discrepancy is the maximum absolute
difference in percentage points. TV compares the complete joint tuple,
including hidden diagnostics rather than only observable acceptance.}
\label{tab:magic-agreement}
\centering\small
\begin{tabular}{@{}lrrrrr@{}}
\toprule
Protocol & Training attempts & Held-out attempts & Max. acceptance difference & Median TV & Max. TV \\
\midrule
Raw T & 2,890,000 & 2,960,000 & 0.00 & 0.000015 & 0.000733 \\
T teleportation & 2,990,000 & 3,010,000 & 0.00 & 0.000820 & 0.003229 \\
15-to-1 & 1,230,000 & 1,220,000 & 0.43 & 0.001575 & 0.012400 \\
CCZ & 1,400,000 & 1,400,000 & 0.20 & 0.000875 & 0.007400 \\
Catalytic CCZ-to-two-T & 1,750,000 & 1,740,000 & 0.00 & 0.003693 & 0.008000 \\
\bottomrule
\end{tabular}
\end{table*}

\paragraph{IBM joint-law comparison.}
The wider IBM dataset spans 32 provider jobs: 16 training-only, one
held-out-only, and 15 containing both categories. These mixed jobs
introduce split dependence. Compiled circuits also differ across some
backend and epoch comparisons.
An exact-circuit postselection/control probe on Marrakesh has one
training and one held-out job, each with 1,024 shots. For categories
(accepted, correct), (accepted, incorrect), (rejected, correct), and
(rejected, incorrect), counts are $(934,20,55,15)$ in training and
$(959,13,35,17)$ held out. Predicting from the training joint
frequencies gives held-out log loss 0.306183 nats/shot and Brier score
0.122347. Products of training marginals give scores 0.336126 and
0.124252, respectively. Both favor retaining dependence
for this pair, without using held-out outcomes to estimate probabilities.
One job per side does not establish a general ranking. The calibration
epoch and Python version differ, so the discrepancy cannot be assigned
solely to hardware drift. This is a physical proxy for joint outcomes,
separate from logical-memory or complete distillation validation.

\subsection{Conditional Memory-Risk Sensitivity}
\label{app:risk-sensitivity}

The reference Z-memory fit gives $p_L=4.44447\times10^{-7}$ per
logical-qubit-round. The timing profile uses physical noise $10^{-3}$
and the hazard fit uses $10^{-4}$, so applying the hazard to timing-derived
exposure is a transferred-reference sensitivity. Its hazard interval
spans $1.05437\times10^{-9}$ to $2.86773\times10^{-6}$, giving
0.28\%--99.95\% conditional failure at uniform reference exposure.

At the fixed reference hazard, joint decoder/factory provisioning changes
uniform conditional memory risk from 69.54\% to 43.39\% and burst risk
from 46.27\% to 29.67\%. These changes follow exposure under unchanged
per-round quality. In the window study, exposure rising from 1.897 to
4.238 million logical-qubit-rounds changes the proxy from 56.96\% to
84.80\%. In the equal-mean variability study, the increase from 2.596 to
5.589 million changes it from 68.46\% to 91.66\%. The broad uncertainty
in absolute probability motivates reporting exposure directly in the
main text. Equation~\ref{eq:survival} states the composition assumption.

\section{Logical Demand Construction}
\label{app:demand-policy}

Section~\ref{sec:demand} defines the matched totals, causal epochs, and
four controlled regimes. The following variants change spacing while
preserving the same work budgets.

\paragraph{Stochastic spacing.}
Draw phase weights and normalize them:
\begin{equation}
 X_k\sim\operatorname{Gamma}(\alpha,1),\quad
 \widetilde g_k=R_p\frac{X_k}{\sum_{j\in p}X_j},\quad k\in p.
 \label{eq:stochastic-gaps}
\end{equation}
Largest-remainder apportionment produces nonnegative integer gaps whose
sum is $R_p$. Shape $\alpha$ controls the raw spacing variability. Normalization and rounding make the final gaps dependent.

\paragraph{Persistent demand.}
A two-state process $Z_k\in\{H,L\}$ instead uses
stationary high-state probability $0<\pi<1$ and persistence
$0\le\eta<1$:
\begin{align}
 \Pr(Z_{k+1}=H\mid Z_k=L)&=\pi(1-\eta),\\
 \Pr(Z_{k+1}=L\mid Z_k=H)&=(1-\pi)(1-\eta).
 \label{eq:demand-persistence}
\end{align}
Mean high and low runs are $[(1-\pi)(1-\eta)]^{-1}$ and
$[\pi(1-\eta)]^{-1}$ demand epochs. State-dependent gap weights,
optionally with Gamma variability, are normalized to the total work
budget. The realized finite-path occupancy is retained. This construction
is an alternative to fixed high/low phase budgets. Its transitions count
demand epochs, not physical QEC rounds.

Demand is generated and saved before execution. Delayed predecessors
shift when later demand is reached. Paired architecture comparisons reuse
the same graph. Demand seeds select different workload realizations. Runtime seeds separately control service and protocol outcomes.

\section{Supplementary Architecture Results}
\label{app:results}

The supplementary comparisons retain the single-job workload, operation
FIFO, physical timing, and profile assumptions of Section~\ref{sec:method}.
U, B, C, and CB denote uniform, burst, clustered, and clustered-burst
demand. All runtimes are job completion times. Synchronization quantiles
are nearest-rank quantiles of completed records, including zero waits.
Concurrent records can describe the same elapsed interval and are treated
as dependent observations. Tables preserve the coordinate grids of the
reported comparisons, with coordinates outside the reported result set
shown as dashes and executions ending
with unfinished work marked $\dagger$. Such executions supply no
completed-runtime or terminal-reliability estimate.

\subsection{Joint Factory and Decoder Provisioning}

\begin{figure*}[tp]
\centering
\includegraphics[width=.96\textwidth]{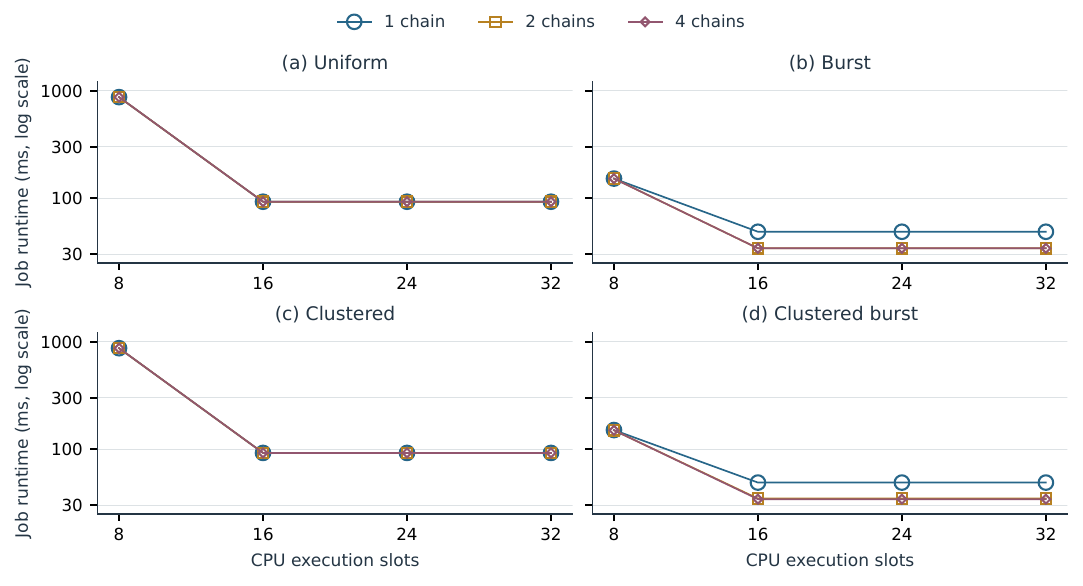}
\caption{Expanded factory/decoder provisioning under (a) uniform,
(b) burst, (c) clustered, and (d) clustered-burst demand. All panels
share a logarithmic runtime scale and retain the reference service
mapping and runtime seed 101. Curves connect discrete completed
coordinates, with open markers of different sizes exposing coincident
values. The grid contains all 48 configurations.}
\label{fig:app-provisioning-extension}
\Description{Four demand-specific panels compare job runtime against CPU
execution slots for one, two, and four factory chains. Additional factories
have little effect at eight slots, while runtime drops sharply at sixteen.}
\end{figure*}

Table~\ref{tab:app-provision} and Figure~\ref{fig:app-provisioning-extension}
expand RQ1 across all four regimes and all 48 factory/CPU combinations.
Uniform demand plateaus at sixteen CPU slots, while burst demand
benefits from a second chain. Clustered bursts retain a small further
benefit from four chains. Every configured producer participates.
Representative one-, two-, and four-chain executions each consume
640 states and leave sixteen in inventory, with respectively eight,
ten, and thirteen rejected attempts. Different producer-specific
realizations explain these counts without changing the acceptance law.

At eight slots, all completed uniform and clustered configurations take
approximately $870$--$871\,\mathrm{ms}$ despite the change in factory count.
Burst demand takes $151.901\,\mathrm{ms}$ with two chains and
$151.745\,\mathrm{ms}$ with four. At sixteen slots, both complete in
$34.029\,\mathrm{ms}$. Thus additional supply has little effect in the
low-capacity setting, while increasing decoder capacity exposes the
faster operating region. Clustered-burst demand shows the same pattern,
with $149.617$ and $149.734\,\mathrm{ms}$ at eight slots for two and
four chains. Its small ordering difference is one realization of
producer-specific outcomes, not evidence of a systematic advantage
for fewer factories. With one chain and eight slots, burst and
clustered-burst demand complete in $151.825$ and $150.430\,\mathrm{ms}$,
respectively.

\begin{table*}[t]
\centering\small
\caption{Complete reference provisioning surface. Entries are job runtime in milliseconds. Each chain includes raw supply, 15-to-1 distillation, and dedicated quantum capacity.}
\label{tab:app-provision}
\begin{tabular}{@{}llrrrr@{}}
\toprule
Demand & Chains & CPU 8 & CPU 16 & CPU 24 & CPU 32 \\
\midrule
U & 1 & 871.051 & 92.451 & 92.451 & 92.451 \\
U & 2 & 870.791 & 92.451 & 92.451 & 92.451 \\
U & 4 & 871.041 & 92.451 & 92.451 & 92.451 \\
B & 1 & 151.825 & 48.561 & 48.561 & 48.561 \\
B & 2 & 151.901 & 34.029 & 34.029 & 34.029 \\
B & 4 & 151.745 & 34.029 & 34.029 & 34.029 \\
C & 1 & 870.448 & 92.123 & 92.123 & 92.123 \\
C & 2 & 870.401 & 92.123 & 92.123 & 92.123 \\
C & 4 & 870.894 & 92.123 & 92.123 & 92.123 \\
CB & 1 & 150.430 & 48.790 & 48.786 & 48.786 \\
CB & 2 & 149.617 & 34.608 & 34.608 & 34.608 \\
CB & 4 & 149.734 & 34.176 & 34.176 & 34.176 \\
\bottomrule
\end{tabular}
\end{table*}

\subsection{How Concurrent Synchronization Waits Overlap}

\begin{figure*}[t]
\centering
\includegraphics[width=.94\textwidth]{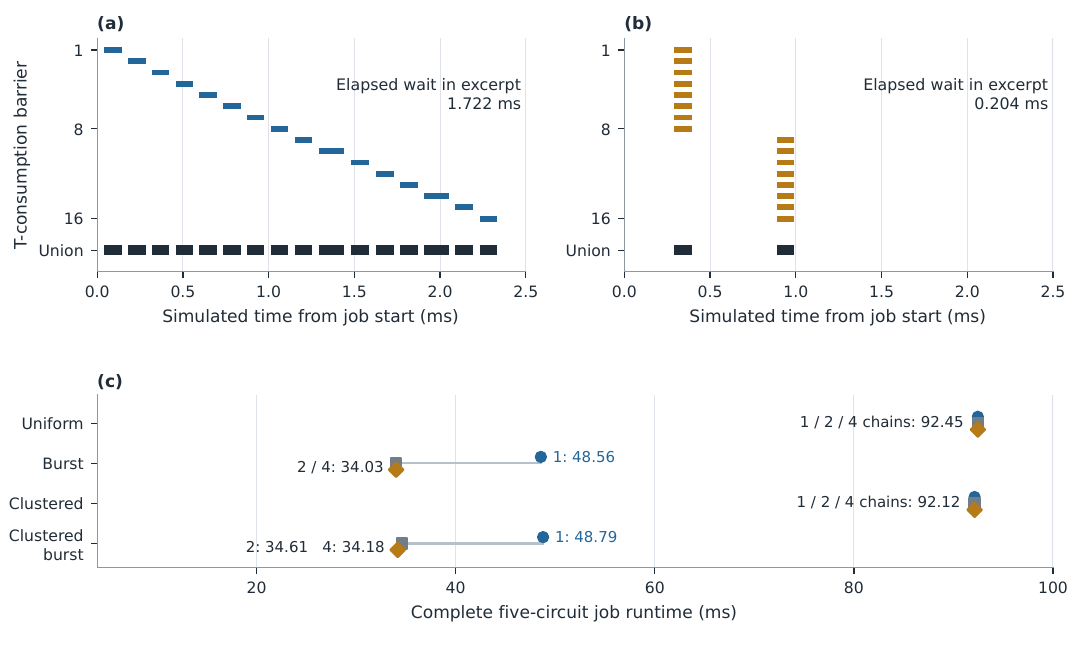}
\caption{Synchronization-wait excerpts and full-job factory provisioning.
Panel (a) shows uniform demand and panel (b) shows burst demand, each with
one factory chain and reference classical capacity. Colored spans
show QEC waits and black strips show their elapsed-time union. The vertical
indices identify consumption barriers. Panel (c) reports full five-circuit
completion times for one, two, and four chains, with small vertical offsets
separating coincident markers. Excerpt waiting and complete-job runtime have
different observation scopes.}
\label{fig:app-qec-waits}
\Description{QEC waiting intervals for uniform and burst demand, followed
by complete-job runtimes for one, two, and four factory chains.}
\end{figure*}

Figure~\ref{fig:app-qec-waits} retains the exact excerpts underlying
Figure~\ref{fig:rq1-scaling}: the first sixteen T consumers and 512 requested
Clifford rounds of the first circuit. Colored intervals run from captured requirement to frontier
resolution. Uniform demand has sixteen separate waits, while two burst
groups share waits. Black strips show elapsed unions, counting concurrent
waits once: 1.722 and 0.204\,ms.
These are QEC waits, excluding any remaining material or resource delay.

\subsection{Storage, Waiting, and Post-Job Replenishment}

Table~\ref{tab:app-storage} reports storage effects for all four regimes.
Equal completion does not imply identical waiting. Four-chain burst
and clustered-burst runs with capacity four each record approximately
$0.484\,\mathrm{ms}$ of summed material waiting. Capacity eight removes
that waiting without shortening the job. These waits overlap other
dependencies, so material waiter time is not an additive runtime component.

The inventory target also changes the work remaining after job completion.
For one-chain burst demand, capacities eight and thirty-two both complete
at $48.561\,\mathrm{ms}$, while the final scheduler horizons are
$48.866\,\mathrm{ms}$ and $50.648\,\mathrm{ms}$. The corresponding
replenishment tails are $0.305\,\mathrm{ms}$ and $2.087\,\mathrm{ms}$.
Counting these tails as application runtime would incorrectly suggest
that the larger buffer slows the job. Uniform runtime is unchanged
across the same storage sweep, although retained inventory and the
number of production attempts increase with the target.

\begin{table*}[t]
\caption{Storage sweep at the reference CPU capacity. Entries are job runtime in milliseconds. Capacity is the common output-store limit in T states.}
\label{tab:app-storage}
\centering
\small
\setlength{\tabcolsep}{5pt}
\begin{tabular}{@{}llrrrr@{}}
\toprule
Demand & Chains & Capacity 4 & Capacity 8 & Capacity 16 & Capacity 32 \\
\midrule
U & 1, 4 & 92.451 & 92.451 & 92.451 & 92.451 \\
B & 1 & 48.968 & 48.561 & 48.561 & 48.561 \\
B & 4 & 34.029 & 34.029 & 34.029 & 34.029 \\
C & 1, 4 & 92.123 & 92.123 & 92.123 & 92.123 \\
CB & 1 & 53.646 & 48.874 & 48.786 & 48.786 \\
CB & 4 & 34.176 & 34.176 & 34.176 & 34.176 \\
\bottomrule
\end{tabular}
\end{table*}

\subsection{Window Concurrency and Overlap}

Section~\ref{sec:rq2} defines the assumed window-service law
(Equation~\ref{eq:window-cost}), its per-new-round work with overlap,
and its calibration limits. The tables below retain the full sweep.

The window sweep in Table~\ref{tab:app-windows} uses this service-cost law.
With uniform demand and three-round windows, raising the window
concurrency limit from eight to twenty-four reduces runtime by
$13.43\times$ at fixed CPU capacity. Limiting outstanding windows constrains how much of
that processor can be used while the protection stream continues.

Increasing overlap introduces repeated decoder work per newly covered
round. With six-round windows, one-third overlap increases uniform
runtime only from $92.451\,\mathrm{ms}$ to $93.873\,\mathrm{ms}$, while
one-half overlap raises it to $517.184\,\mathrm{ms}$
(Table~\ref{tab:app-overlap}). The latter has p99 synchronization wait
$820.3\us$ and 12.431 million logical-qubit-rounds of exposure.
The reported decoder utilization increases from approximately 44.6\%
without overlap to 66.9\% at one-third overlap and 74.0\% at one-half
overlap. Thus the additional work can move the system into a different
backlog regime. With six rounds and one-half overlap, burst and
clustered demand complete in $89.750$ and $517.260\,\mathrm{ms}$.
With three-round history and eight outstanding windows, they take
$161.380$ and $879.542\,\mathrm{ms}$.

For burst demand with one chain and 24 concurrent windows, changing
window size mainly changes the balance between QEC and material waits.
At three and twelve rounds, QEC wait unions are respectively
$5.371\,\mathrm{ms}$ and $15.933\,\mathrm{ms}$, while summed material
waits fall from $41.202\,\mathrm{ms}$ to $10.480\,\mathrm{ms}$.
These distinct measures should not be added. Windowing and overlap
quality benefits require condition-matched error profiles and are
outside this timing sensitivity.

\begin{table}[t]
\caption{Window size and outstanding-window capacity at the reference CPU capacity and one production chain. Entries are job runtime in milliseconds.}
\label{tab:app-windows}
\centering
\small
\setlength{\tabcolsep}{4pt}
\begin{tabular}{@{}llrr@{}}
\toprule
Demand & Rounds & 8 windows & 24 windows \\
\midrule
U & 3 & 880.258 & 65.525 \\
U & 9 & 733.233 & 119.534 \\
U & 12 & 733.814 & 146.617 \\
B & 3 & 161.380 & 48.480 \\
B & 9 & 131.969 & 48.649 \\
B & 12 & 130.704 & 48.719 \\
C & 3 & 879.542 & 65.668 \\
C & 9 & 732.849 & 119.253 \\
C & 12 & 732.747 & 146.390 \\
CB & 3 & 160.036 & 48.720 \\
CB & 9 & 131.247 & 49.041 \\
CB & 12 & 129.972 & 49.749 \\
\bottomrule
\end{tabular}
\end{table}

\begin{table}[t]
\caption{Overlap sweep with 24 concurrent windows, reference CPU capacity, and one production chain. Entries are job runtime in milliseconds. Overlap fractions are relative to window size.}
\label{tab:app-overlap}
\centering
\small
\setlength{\tabcolsep}{4pt}
\begin{tabular}{@{}llrrr@{}}
\toprule
Demand & Rounds & Zero & One-third & One-half \\
\midrule
U & 6 & 92.451 & 93.873 & 517.184 \\
U & 12 & 146.617 & 148.990 & 507.402 \\
B & 6 & 48.561 & 48.565 & 89.750 \\
B & 12 & 48.719 & 48.739 & 80.563 \\
C & 6 & 92.123 & 93.780 & 517.260 \\
C & 12 & 146.390 & 148.437 & 507.548 \\
CB & 6 & 48.786 & 48.800 & 90.523 \\
CB & 12 & 49.749 & 49.766 & 81.211 \\
\bottomrule
\end{tabular}
\end{table}

\subsection{Predecoder Fallback and Classical Headroom}

Table~\ref{tab:app-predecode} includes both demand regimes at sixteen
and thirty-two CPU slots, which give identical completion times.
The assumed $4.731\us$ GPU service retains the common link costs,
with fallback executing all four original CPU calls. Zero fallback
is an idealized fast path with uncalibrated accuracy.

For uniform demand, 1\% configured fallback realizes 68 fallbacks among
7,290 windows. The QEC wait union increases from $7.438$ to
$13.856\,\mathrm{ms}$ with nearly unchanged completion. At 10\%,
the longer protection interval itself generates more windows, adding
work beyond the higher frequency of the expensive path.

Clustered bursts remain close to $48.6$--$48.8\,\mathrm{ms}$ across the
fallback sweep. The zero-fallback case has a QEC wait union of
approximately $1.000\,\mathrm{ms}$ and $78.859\,\mathrm{ms}$ of summed
material waiting. At 30\% fallback those quantities become
$8.247\,\mathrm{ms}$ and $48.107\,\mathrm{ms}$. Faster decoding exposes
the material-supply dependency earlier, explaining why stage
acceleration can primarily redistribute waiting.

\begin{table}[t]
\caption{Assumed predecoder sweep. Runtime is in milliseconds and synchronization quantiles in microseconds. Results apply to both tested CPU capacities, sixteen and thirty-two execution slots.}
\label{tab:app-predecode}
\centering
\small
\setlength{\tabcolsep}{4pt}
\begin{tabular}{@{}llrrr@{}}
\toprule
Demand & Fallback & Runtime & p95 & p99 \\
\midrule
U & 0\% & 48.168 & 12.8 & 12.8 \\
U & 1\% & 48.169 & 91.3 & 105.4 \\
U & 10\% & 75.342 & 111.4 & 123.4 \\
U & 30\% & 90.514 & 115.7 & 133.3 \\
CB & 0\% & 48.606 & 12.8 & 12.8 \\
CB & 1\% & 48.605 & 90.0 & 101.3 \\
CB & 10\% & 48.769 & 108.4 & 117.0 \\
CB & 30\% & 48.794 & 117.0 & 126.7 \\
\bottomrule
\end{tabular}
\end{table}

\subsection{Latency Distributions and Capacity}

Table~\ref{tab:app-tailcapacity} expands the equal-mean comparison across
CPU capacity. Constant service yields no completion improvement from
sixteen to thirty-two execution slots in these cases. Tail-heavy service gives a
modest improvement from sixteen to twenty-four slots, after which
completion plateaus. In the uniform case, constant and tail-heavy
service at the reference capacity use approximately 44.6\% and 44.5\% reported decoder
utilization, respectively, despite the $2.16\times$ runtime difference.

Exact records support descriptive latency-budget comparisons without
changing the execution. For uniform demand, none of the constant or
empirical-reference synchronization records exceed $200\us$, while
81.85\% of the tail-heavy records do. This is a post hoc exceedance
fraction for the observed trace. No deadline was installed in the
scheduler, and the records are not independent trials of a deadline
failure probability.

\begin{table*}[t]
\caption{Constant and equal-expected-mean tail-heavy service by CPU capacity. Runtime is in milliseconds and p99 in microseconds. CPU capacities 24 and 32 have identical runtimes. The p99 column uses 32 slots.}
\label{tab:app-tailcapacity}
\centering
\small
\setlength{\tabcolsep}{5pt}
\begin{tabular}{@{}llrrr@{}}
\toprule
Demand & Service & CPU 16 & CPU 24/32 & p99 \\
\midrule
U & Constant & 89.753 & 89.753 & 101.2 \\
U & Tail & 198.863 & 193.555 & 407.2 \\
B & Constant & 48.557 & 48.557 & 102.7 \\
B & Tail & 48.984 & 48.973 & 401.7 \\
C & Constant & 90.566 & 90.566 & 102.7 \\
C & Tail & 195.583 & 191.253 & 409.2 \\
CB & Constant & 48.784 & 48.784 & 102.7 \\
CB & Tail & 51.934 & 51.875 & 404.0 \\
\bottomrule
\end{tabular}
\end{table*}

\subsection{Stochastic Demand Robustness}

Table~\ref{tab:app-demand-robustness} varies within-phase spacing or
phase persistence while preserving total Clifford work and T demand.
The Gamma variants use shapes 0.5 and 2 with exact-budget
apportionment. The Markov variants use stationary high-state probability
0.5 and persistence parameters 0.5 and 0.875. Finite realized occupancies
and circuit offsets remain part of each frozen workload.

Singleton clustered demand spans only $92.113$--$92.477\,\mathrm{ms}$
across these variants. Clustered-burst demand with Gamma shape 0.5
instead reaches $52.205\,\mathrm{ms}$, 7.0\% above the regular reference,
and raises summed material waiting from $46.217\,\mathrm{ms}$ to
$60.592\,\mathrm{ms}$. Its logical-qubit-round exposure increases by
7.1\%. The tested Markov variants complete close to $48.56\,\mathrm{ms}$.
The combination of burst width and temporal concentration therefore
matters more than spacing variability alone in this workload. These
single-realization comparisons do not define a universal ranking of
stochastic demand families.

\begin{table*}[t]
\caption{Frozen demand variants at the reference CPU capacity and one chain. Runtime and summed material waiting are in milliseconds, p99 is in microseconds, and exposure is in millions of logical-qubit-rounds. Conditional failure uses the controlled reference hazard.}
\label{tab:app-demand-robustness}
\centering
\small
\setlength{\tabcolsep}{5pt}
\begin{tabular}{@{}llrrrrr@{}}
\toprule
Demand & Spacing & Runtime & Material wait & p99 & Exposure & Failure (\%) \\
\midrule
C & Regular & 92.123 & 0.000 & 135.8 & 2.665 & 69.41 \\
C & Gamma $\alpha=0.5$ & 92.286 & 0.000 & 132.3 & 2.669 & 69.47 \\
C & Gamma $\alpha=2$ & 92.113 & 0.000 & 131.3 & 2.665 & 69.41 \\
C & Markov $\eta=0.5$ & 92.464 & 0.000 & 140.0 & 2.675 & 69.55 \\
C & Markov $\eta=0.875$ & 92.477 & 0.000 & 136.7 & 2.675 & 69.54 \\
CB & Regular & 48.786 & 46.217 & 137.3 & 1.404 & 46.43 \\
CB & Gamma $\alpha=0.5$ & 52.205 & 60.592 & 145.7 & 1.504 & 48.74 \\
CB & Gamma $\alpha=2$ & 48.945 & 46.265 & 131.4 & 1.409 & 46.53 \\
CB & Markov $\eta=0.5$ & 48.573 & 38.696 & 152.7 & 1.397 & 46.24 \\
CB & Markov $\eta=0.875$ & 48.563 & 31.657 & 130.5 & 1.398 & 46.27 \\
\bottomrule
\end{tabular}
\end{table*}

\subsection{Factory Dependence and Classical-Control Sensitivity}

The factory-dependence comparison retains the acceptance and output
marginals, but changes how classical control duration is associated with
the outcome. Control takes $300\,\mathrm{ns}$ with probability 0.9 and
$3{,}300\,\mathrm{ns}$ with probability 0.1, retaining a
$600\,\mathrm{ns}$ mean. The correlated law aligns slow service and
rejection as far as the marginals permit. The factorized law separates
those fields. Physical quantum stages remain unchanged, and each attempt
uses its selected duration without padding to the longest branch.

All four matched comparisons give identical runtime and exposure.
For one-chain clustered bursts, summed material waiting differs only
from $46.105$ to $46.111\,\mathrm{ms}$. The dependence change does not
reach the application critical path in these configurations. Its system
impact depends on the correlated stage's position relative to other
dependencies.

A separate sweep changes classical frame/apply service, keeping physical
quantum operations intact (Table~\ref{tab:app-control}). Increasing that
service from 100 to $1{,}000\,\mathrm{ns}$ raises uniform runtime from
$89.900\,\mathrm{ms}$ to $98.690\,\mathrm{ms}$, approximately 9.8\%.
The conditional memory-failure proxy increases from 68.52\% to 71.87\%.
Burst completion is less sensitive because production continues to
constrain progress. The explicit resource-bound control path therefore
remains relevant even when its individual operations are short.

\begin{table*}[t]
\caption{Classical frame/apply sensitivity with one chain and the reference CPU capacity. Runtime is in milliseconds. Failure is the aggregate conditional memory proxy at the fixed reference hazard.}
\label{tab:app-control}
\centering
\small
\setlength{\tabcolsep}{5pt}
\begin{tabular}{@{}lrrrrr@{}}
\toprule
Demand & 100 ns & 600 ns & 1{,}000 ns & Failure at 100 ns (\%) & Failure at 1{,}000 ns (\%) \\
\midrule
U & 89.900 & 92.451 & 98.690 & 68.52 & 71.87 \\
B & 48.555 & 48.561 & 48.589 & 46.28 & 46.25 \\
C & 90.184 & 92.123 & 98.769 & 68.63 & 71.91 \\
CB & 48.786 & 48.786 & 48.832 & 46.44 & 46.42 \\
\bottomrule
\end{tabular}
\end{table*}

\subsection{Future Decoder Speed and Reliability Sensitivity}

Decoder speed scales CPU service by $1/s$, preserving the $1.1\us$ QEC
cadence, physical protocol service, topology, and per-round reliability
input. Table~\ref{tab:app-speed} includes slower and faster service for
all four regimes and both chain counts. Half-speed decoding raises
uniform runtime to about $582\,\mathrm{ms}$. At twice reference speed,
uniform completion is equal for one and four chains. At fourfold speed,
additional chains reduce the remaining material limitation.
For regular bursts, four chains expose decoder improvements that one
chain largely hides: runtime falls from $34.029\,\mathrm{ms}$ at
reference speed to $27.372\,\mathrm{ms}$ at fourfold speed.
Clustered bursts show the same interaction, reaching
$27.395\,\mathrm{ms}$ with four chains and fourfold decoding.

Applying assumed hazards to the same measured exposure isolates
reliability sensitivity without changing execution. For
uniform reference execution, changing the hazard from $10^{-7}$ to
$10^{-6}$ changes conditional failure from 23.47\% to 93.11\%.
Jointly increasing decoder speed and factory capacity reduces those
values to 12.01\% and 72.20\%, respectively. Absolute values are sensitive
to the hazard, while the exposure ordering is preserved for any common
positive hazard:
\begin{equation}
  \frac{\partial}{\partial E}\bigl[1-(1-p_L)^E\bigr]
  =-(1-p_L)^E\log(1-p_L)>0,\quad 0<p_L<1.
\end{equation}
Changing decoder quality or code distance can change $p_L$ as well as
$E$, requiring a joint comparison instead of this fixed-hazard ordering.

The reference $p_L=4.44447\times10^{-7}$ comes from the independent
round-parity fit, with retained 95\% bounds
$[1.05437\times10^{-9},\,2.86773\times10^{-6}]$.
Training uses six- and twelve-round coordinates with two logical failures
among 500,000 samples at each, plus a held-out round-count observation.
The interval maps to 0.28\%--99.95\% conditional failure at uniform
reference exposure. This uncertainty is conditional on the fit and
does not resolve the physical-noise mismatch between memory and timing.

The factory profile supplies a separate quality diagnostic. Its accepted
output atom has zero observed residual density in the finite empirical
population, while rejected outcomes carry nonzero diagnostics. This
does not establish zero physical accepted-output error. The present
architecture sweep does not vary accepted-output quality, and no
combined algorithm-success estimate is formed from this diagnostic
and the data-memory proxy. Physical-qubit-time reported for protection
also excludes the cost of additional factory hardware.

\begin{table*}[tp]
\caption{Decoder-speed sensitivity at the reference CPU capacity. Runtime is in milliseconds. The final two columns give exposure in millions of logical-qubit-rounds and aggregate conditional failure at speed four and the reference hazard.}
\label{tab:app-speed}
\centering
\small
\setlength{\tabcolsep}{5pt}
\begin{tabular}{@{}llrrrrrr@{}}
\toprule
Demand & Chains & $s=0.5$ & $s=1$ & $s=2$ & $s=4$ & Exposure, $s=4$ & Failure, $s=4$ (\%) \\
\midrule
U & 1 & 581.653 & 92.451 & 60.222 & 48.208 & 1.398 & 46.28 \\
U & 4 & 581.653 & 92.451 & 60.222 & 44.153 & 1.280 & 43.39 \\
B & 1 & 99.119 & 48.561 & 48.465 & 48.418 & 1.404 & 46.42 \\
B & 4 & 99.111 & 34.029 & 29.570 & 27.372 & 0.792 & 29.67 \\
C & 1 & 582.037 & 92.123 & 60.318 & 49.686 & 1.441 & 47.30 \\
C & 4 & 582.037 & 92.123 & 59.954 & 44.117 & 1.279 & 43.36 \\
CB & 1 & 99.187 & 48.786 & 48.685 & 48.642 & 1.411 & 46.58 \\
CB & 4 & 98.105 & 34.176 & 29.600 & 27.395 & 0.793 & 29.69 \\
\bottomrule
\end{tabular}
\end{table*}

\subsection{Continuous-Inventory Repetition}

The repetition comparison doubles shots per circuit with the same
five demand graphs and continuous inventory contract. Each shot retains
initialization and readout. Two-shot jobs consume 1,280 T states and
resolve 1,300 synchronization records. Mean time per circuit-shot
execution changes by less than 0.65\% in every regime, preserving the
singleton/burst ordering. For uniform, burst, clustered, and clustered-burst
demand, respectively, the means change from $18.490$ to $18.426$,
$9.712$ to $9.668$, $18.425$ to $18.468$, and $9.757$ to
$9.694\,\mathrm{ms}$. These divide job totals, including outer stages,
by five or ten executions.

Shared factory state and workload realizations make these dependent
repetitions. Reliability comparisons across shot counts use exposure
per execution or a declared normalized proxy, since aggregate survival
covers different amounts of work.

\section{Code- and Decoder-Dependent Studies}
\label{app:additional-model}

\subsection{Matched Family Results}

The family study uses the workload and service mapping in
Section~\ref{sec:method}. A measurement ancilla is assumed per retained
stabilizer check. Geometry determines full syndrome payload, while
code-dependent initialization, injection, and readout each take $d$
rounds. Factory technology remains fixed at the reference $d=3$.
Color/HGP matrix-sector noise and surface-code circuit-noise evidence
have distinct domains. These settings compare architectural service
envelopes, with intrinsic accuracy left uncalibrated across families.
A common positive hazard preserves exposure ordering, while
family-specific LER rankings and algorithm success require additional
calibration and error composition.

\begin{table*}[t]
\centering\small
\caption{Matched code/decoder provisioning studies. Runtime is in milliseconds. Each row retains its workload and service mapping while changing one to two complete factory chains. Burst reductions are relative to one chain. HGP data-qubit counts are 72, 200, and 392. Dashes identify the four incomplete uniform executions, which contribute no completion estimates. All 56 completed runs consume 640 T states and resolve 650 synchronization records.}
\label{tab:family-results}
\begin{tabular}{@{}lrrrrrrr@{}}
\toprule
Code family & $d$ & Uniform, 1 & Uniform, 2 & Burst, 1 & Burst, 2 & Runtime reduction & Exposure reduction \\
\midrule
Rotated surface & 3 & 126.222 & 126.222 & 48.669 & 38.821 & 20.24\% & 20.56\% \\
Rotated surface & 5 & 152.657 & 152.657 & 48.749 & 43.647 & 10.47\% & 10.68\% \\
Rotated surface & 7 & 213.191 & 213.191 & 53.419 & 53.419 & 0.00\% & 0.00\% \\
Unrotated surface & 3 & 132.617 & 132.617 & 48.685 & 39.513 & 18.84\% & 19.16\% \\
Unrotated surface & 5 & 181.953 & 181.953 & 48.838 & 47.608 & 2.52\% & 2.58\% \\
Unrotated surface & 7 & 294.957 & 294.957 & 64.583 & 64.583 & 0.00\% & 0.00\% \\
Toric surface & 3 & 81.121 & 81.121 & 48.528 & 32.311 & 33.42\% & 33.72\% \\
Toric surface & 5 & 117.605 & 117.605 & 48.641 & 38.701 & 20.44\% & 20.73\% \\
Toric surface & 7 & 169.839 & 169.839 & 48.804 & 47.469 & 2.74\% & 2.81\% \\
Color code & 3 & 578.813 & 578.813 & 101.487 & 101.487 & 0.00\% & 0.00\% \\
Color code & 5 & 539.485 & 539.485 & 97.270 & 97.270 & 0.00\% & 0.00\% \\
Color code & 7 & 570.208 & 570.208 & 102.667 & 102.667 & 0.00\% & 0.00\% \\
HGP & 6 & 284.931 & 284.931 & 62.636 & 62.636 & 0.00\% & 0.00\% \\
HGP & 10 & -- & -- & 112.120 & 112.120 & 0.00\% & 0.00\% \\
HGP & 14 & -- & -- & 162.777 & 162.777 & 0.00\% & 0.00\% \\
\bottomrule
\end{tabular}
\end{table*}

\subsection{Color-Code and HGP Window Sensitivity}

Color-code comparisons at $d=3,5$ and HGP at $d=6$ retain the family-study
CPU capacity, one chain, and runtime seed 101. Each code-block call
uses Equation~\ref{eq:window-cost}, with $W_0=2s$ for construction
size $s$. Doubling history without overlap reduces service per newly
resolved round by 10\%, yet increases runtime and protected exposure.
Adding half-history overlap approximately preserves completion at
this capacity setting. Decoder accuracy remains uncalibrated.

Table~\ref{tab:color-window-extension} includes uniform and burst
demand. Under uniform demand, doubled history increases runtime
$1.77\times$ for color $d=3$, $1.76\times$ for color $d=5$, and
$1.72\times$ for HGP $d=6$. The HGP burst increase is $1.46\times$.
All these configurations record zero material waiting, linking the
effects to information readiness under the specified service mapping.

\begin{table*}[t]
\centering\small
\caption{Matched window comparisons with one chain and 256 CPU slots.
Runtime is in milliseconds and exposure in millions of logical-qubit-rounds.
U and B denote uniform and burst demand. The first row for each code is
its family-study control. All twelve extended-window runs shown complete.}
\label{tab:color-window-extension}
\begin{tabular}{@{}lrrrrrrr@{}}
\toprule
Code & $d$ & History & Overlap & Runtime U & Runtime B & Exposure U & Exposure B \\
\midrule
Color & 3 & 6 & 0 & 578.813 & 101.487 & 16.714 & 2.828 \\
Color & 3 & 12 & 0 & 1022.795 & 162.967 & 29.531 & 4.518 \\
Color & 3 & 12 & 6 & 1024.909 & 163.048 & 29.592 & 4.520 \\
Color & 5 & 10 & 0 & 539.485 & 97.270 & 15.579 & 2.715 \\
Color & 5 & 20 & 0 & 950.604 & 154.183 & 27.448 & 4.279 \\
Color & 5 & 20 & 10 & 952.941 & 154.214 & 27.515 & 4.280 \\
HGP & 6 & 6 & 0 & 284.931 & 62.636 & 8.231 & 1.764 \\
HGP & 6 & 12 & 0 & 490.854 & 91.156 & 14.176 & 2.548 \\
HGP & 6 & 12 & 6 & 492.963 & 91.174 & 14.237 & 2.549 \\
\bottomrule
\end{tabular}
\end{table*}

\subsection{Repeated-Seed Decoder Variability}

\begin{figure*}[tp]
\centering
\includegraphics[width=.96\textwidth]{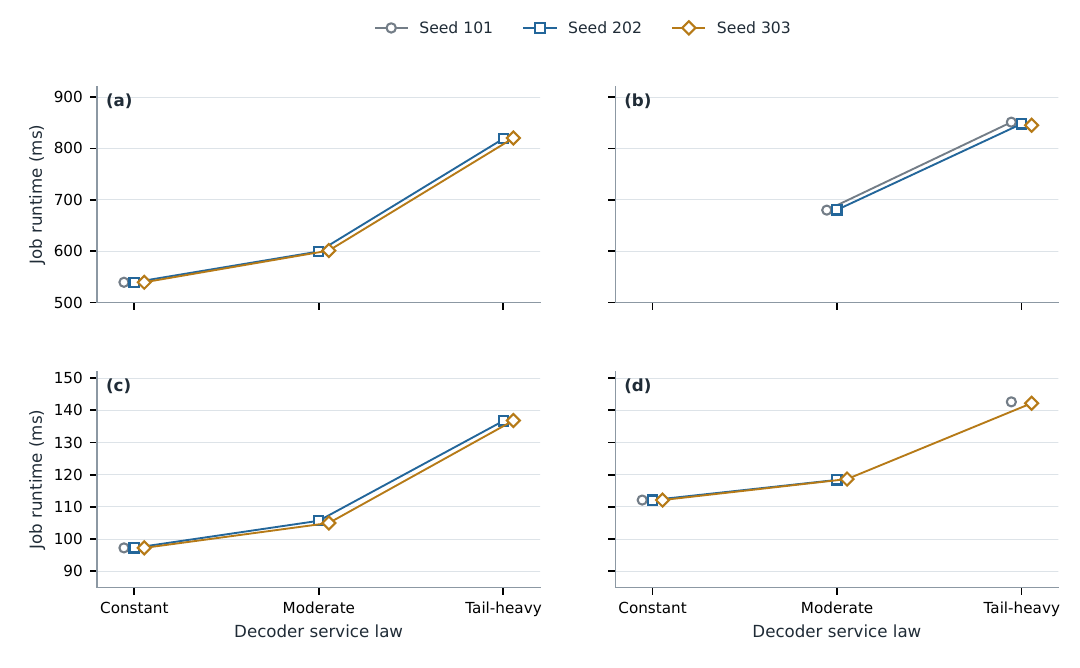}
\caption{Individual-seed runtime under equal-mean decoder service laws.
Panels (a) and (b) show uniform demand for color $d=5$ and HGP $d=10$,
respectively. Panels (c) and (d) show burst demand in the same family
order. Runtime scales are shared within each demand row. Horizontal
offsets separate seed markers at the same categorical law. Lines
guide the eye between adjacent completed configurations, with no
interpolation through unavailable outcomes. No constant-service
uniform HGP execution completes. Exposure and missing-outcome status
are retained in Table~\ref{tab:app-family-seeds}.}
\label{fig:app-family-variability}
\Description{Four panels compare constant, moderate, and tail-heavy
decoder service across individual runtime seeds. Completed color-code
comparisons consistently show higher runtime with heavier tails.
HGP comparisons have incomplete seed coverage.}
\end{figure*}

This supplement extends the seed-101 studies in
Section~\ref{sec:method} with runtime seeds 202 and 303, retaining
the same saved logical demand.
Color code at $d=5$ and the 200-data-qubit HGP code ($d=10$) retain
ten-round history, zero overlap, one chain, and the family-study service
mapping. The constant law has service $m$. The moderate law selects
$0.5m$ or $1.5m$ with equal probability, while the tail-heavy law
selects $0.5m$ with probability $0.9$ and $5.5m$ otherwise. All three
have equal expected mean. Figure~\ref{fig:app-family-variability}
shows individual runtime seeds and Table~\ref{tab:app-family-seeds}
retains runtime, protected exposure, and the elapsed QEC-wait union.
The family mapping is separate from the reference mapping in
Figure~\ref{fig:rq3-results}, so their observations are not pooled.

For color code, both seeds with complete three-law comparisons
preserve the same ordering. Under uniform demand, tail-heavy service
increases runtime by $51.8$--$52.0\%$ and exposure by
$51.8$--$52.0\%$ relative to constant service. Under burst demand,
runtime rises $40.5$--$40.6\%$ and exposure rises $39.3$--$40.0\%$.
Moderate variability produces intermediate runtime and exposure in
both regimes. Thus the sensitivity to service shape persists across
these realizations, including burst demand where the reference
factory-limited configuration largely hides the runtime effect.

For HGP burst demand, matched constant/tail-heavy pairs at seeds 101
and 303 show runtime increases of $26.8$--$27.2\%$ and exposure
increases of $26.1$--$26.3\%$. At seed 101, peak backlog rises from
811 to 1,475 rounds ($81.9\%$), preserving the earlier matched
comparison. For uniform demand, the constant controls end unfinished.
The completed moderate/tail-heavy pair at seed 101 instead gives
$25.2\%$ more runtime and $25.1\%$ more exposure under the heavier
tail. Seed 202 retains this ordering. The missing moderate outcome
at seed 303 precludes a third matched uniform comparison.

All 26 completed family-variability executions have zero recorded
material waiting. The observations therefore connect service shape
to information readiness and protection overhead under these
architectural settings. Points show individual executions, with
absent outcomes distinguished in the table. The small seed count
supports a consistency check rather than precise confidence intervals.
The service laws are synthetic and family exposure remains
uncalibrated, with no family-specific LER or algorithm-success claim.

\begin{table*}[t]
\centering\footnotesize
\setlength{\tabcolsep}{2.7pt}
\caption{Individual-seed variability results. Time is job runtime in milliseconds, $E$ is exposure in millions of logical-qubit-rounds, and Wait is the QEC-wait union in milliseconds. U and B denote uniform and burst demand. All coordinates use ten-round history, zero overlap, one chain, and 256 CPU slots. Dashes denote unavailable remote results and daggers denote incomplete executions.}
\label{tab:hgp-tail-extension}
\label{tab:hgp-additional}
\label{tab:app-family-seeds}
\begin{tabular}{@{}lllrrrrrrrrr@{}}
\toprule
& & & \multicolumn{3}{c}{Seed 101} & \multicolumn{3}{c}{Seed 202} & \multicolumn{3}{c}{Seed 303} \\
\cmidrule(lr){4-6}\cmidrule(lr){7-9}\cmidrule(l){10-12}
Code & Demand & Law & Time & $E$ & Wait & Time & $E$ & Wait & Time & $E$ & Wait \\
\midrule
Color $d=5$ & U & Constant & 539.485 & 15.579 & 509.837 & 539.485 & 15.579 & 509.837 & 539.485 & 15.579 & 509.837 \\
Color $d=5$ & U & Moderate & \multicolumn{3}{c}{---} & 599.358 & 17.303 & 569.046 & 601.217 & 17.361 & 571.032 \\
Color $d=5$ & U & Tail-heavy & \multicolumn{3}{c}{---} & 818.784 & 23.642 & 786.940 & 820.176 & 23.680 & 788.249 \\
Color $d=5$ & B & Constant & 97.270 & 2.715 & 67.265 & 97.270 & 2.715 & 67.265 & 97.270 & 2.715 & 67.265 \\
Color $d=5$ & B & Moderate & \multicolumn{3}{c}{---} & 105.678 & 2.949 & 75.301 & 104.989 & 2.924 & 74.459 \\
Color $d=5$ & B & Tail-heavy & \multicolumn{3}{c}{---} & 136.681 & 3.782 & 103.954 & 136.804 & 3.801 & 104.615 \\
\midrule
HGP $d=10$ & U & Constant & \multicolumn{3}{c}{$\dagger$} & \multicolumn{3}{c}{$\dagger$} & \multicolumn{3}{c}{$\dagger$} \\
HGP $d=10$ & U & Moderate & 679.912 & 18.670 & 642.900 & 680.372 & 18.682 & 643.320 & \multicolumn{3}{c}{$\dagger$} \\
HGP $d=10$ & U & Tail-heavy & 851.112 & 23.359 & 812.741 & 847.784 & 23.265 & 809.355 & 845.229 & 23.200 & 806.963 \\
HGP $d=10$ & B & Constant & 112.120 & 3.007 & 75.472 & 112.120 & 3.007 & 75.472 & 112.120 & 3.007 & 75.472 \\
HGP $d=10$ & B & Moderate & \multicolumn{3}{c}{$\dagger$} & 118.372 & 3.170 & 81.363 & 118.635 & 3.177 & 81.608 \\
HGP $d=10$ & B & Tail-heavy & 142.614 & 3.797 & 104.012 & \multicolumn{3}{c}{$\dagger$} & 142.186 & 3.790 & 103.757 \\
\bottomrule
\end{tabular}
\end{table*}

\FloatBarrier

\clearpage
\section{Complete-Application Supporting Experiments}
\label{app:application-sweeps}

This supplement retains absolute joint-provisioning times and the detailed
window, fallback, storage, controller, dependence, and repetition studies.
Each table identifies its matched mapping.

\begin{table*}[t]
\centering\small
\caption{Median application completion (ms). F: complete chains. D: decoder speed factor. Each entry summarizes three runtime seeds.}
\label{tab:application-joint}
\begin{tabular}{@{}rrrr@{}}
\toprule
F & D & Multiplier & Ising \\
\midrule
1 & 1 & 52.992 & 536.132 \\
1 & 4 & 52.750 & 536.069 \\
2 & 1 & 28.733 & 278.642 \\
2 & 4 & 26.620 & 268.694 \\
4 & 1 & 21.453 & 214.643 \\
4 & 4 & 13.441 & 135.856 \\
\bottomrule
\end{tabular}
\end{table*}

\subsection{Absolute Completion Across Family Mappings}
\begin{figure*}[t]
\centering
\includegraphics[width=.94\textwidth]{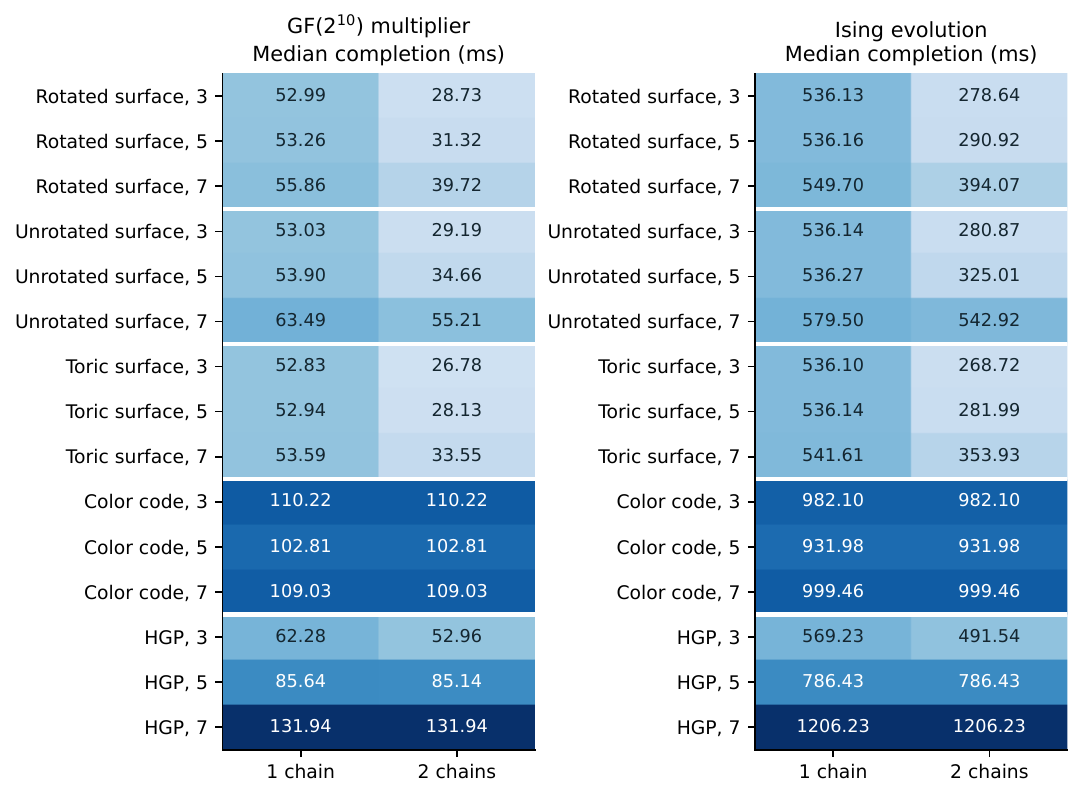}
\caption{Complete absolute-time matrix behind the provisioning reductions
in Figure~\ref{fig:application-family-matrix}. Cells show median completion
in milliseconds over three seeds. Compare one and two chains within a
family and size. Each panel uses its own scale.}
\label{fig:application-family-absolute}
\Description{Absolute completion times for both applications, five code
mappings, three sizes, and one or two factory chains.}
\end{figure*}

\subsection{Window and Family-Window Details}
\begin{table*}[t]
\centering\small
\caption{Every retained reference history/overlap/outstanding-capacity setting. M: multiplier. I: Ising. Values are three-seed medians in milliseconds. Service sums count concurrent decoder work. QEC waiting is an elapsed union. Neither is a decomposition of completion. The six-round, eight-outstanding setting was not run.}
\label{tab:application-window-work}
\begin{tabular}{@{}lrrrrrr@{}}
\toprule
$W/O/P$ & M: time & M: service & M: QEC wait & I: time & I: service & I: QEC wait \\
\midrule
3/0/8 & 262.562 & 2075.265 & 258.816 & 536.095 & 4124.345 & 210.099 \\
9/0/8 & 220.359 & 1752.788 & 216.842 & 536.169 & 3208.100 & 482.571 \\
12/0/8 & 221.037 & 1759.502 & 218.129 & 543.449 & 3135.418 & 530.672 \\
3/0/24 & 86.441 & 2048.183 & 85.038 & 536.095 & 4124.345 & 210.099 \\
6/0/24 & 52.992 & 1129.275 & 47.057 & 536.132 & 3437.081 & 347.968 \\
9/0/24 & 53.416 & 1061.428 & 51.602 & 536.169 & 3208.100 & 482.571 \\
12/0/24 & 55.303 & 1058.748 & 54.502 & 543.449 & 3135.418 & 530.672 \\
6/2/24 & 121.564 & 2894.137 & 119.763 & 536.132 & 5155.621 & 351.636 \\
6/3/24 & 157.141 & 3741.446 & 154.826 & 536.134 & 6874.204 & 353.276 \\
12/4/24 & 126.129 & 3009.689 & 124.399 & 545.796 & 4723.388 & 533.025 \\
12/6/24 & 158.445 & 3783.311 & 156.205 & 545.802 & 6297.952 & 533.148 \\
\bottomrule
\end{tabular}
\end{table*}

\paragraph{Longer family windows can consume much more work.}
Table~\ref{tab:application-family-windows} adds the remaining family
window configurations. At color-code coordinate five, doubling history
from ten to twenty rounds increases multiplier completion from 102.807
to 183.559\,ms. Adding ten-round overlap then changes completion only
to 183.888\,ms, although summed decoder service approximately doubles,
from 10,911.817 to 21,863.155\,ms. Thus a small endpoint difference
can hide a large difference in resource work even outside the
reference supply-limited setting. The table separates the effect of
longer history from the additional cost of overlapping that history.

\begin{table*}[t]
\centering\small
\caption{Family-window sensitivity at one chain, 256 CPU slots and 256 outstanding windows. M: multiplier. I: Ising. Values are three-seed medians in ms. Each size uses its own retained mean and a service factor $0.2+0.8W/(2s)$. Doubling history and introducing overlap are separate interventions.}
\label{tab:application-family-windows}
\begin{tabular}{@{}lrlrrrr@{}}
\toprule
Mapping & $s$ & $W/O$ & M: time & M: service & I: time & I: service \\
\midrule
Color code & 3 & 6/0 & 110.223 & 13165.000 & 982.101 & 35439.491 \\
Color code & 3 & 12/0 & 197.284 & 21207.093 & 1756.300 & 57039.175 \\
Color code & 3 & 12/6 & 197.517 & 42462.783 & 1756.634 & 114099.791 \\
Color code & 5 & 10/0 & 102.807 & 6791.068 & 931.975 & 18597.822 \\
Color code & 5 & 20/0 & 183.559 & 10911.817 & 1638.981 & 29435.822 \\
Color code & 5 & 20/10 & 183.888 & 21863.155 & 1662.295 & 59708.703 \\
HGP & 3 & 6/0 & 62.280 & 3434.305 & 569.229 & 9472.331 \\
HGP & 3 & 12/0 & 93.393 & 4630.214 & 858.646 & 12859.078 \\
HGP & 3 & 12/6 & 93.527 & 9272.952 & 860.103 & 25761.854 \\
HGP & 5 & 10/0 & 85.639 & 4575.206 & 786.427 & 12689.849 \\
HGP & 5 & 20/0 & 150.521 & 7236.176 & 1378.163 & 20014.475 \\
HGP & 5 & 20/10 & 151.084 & 14526.678 & 1385.398 & 40238.910 \\
\bottomrule
\end{tabular}
\end{table*}

\subsection{Fallback, Storage, Control, Dependence and Repetition}
\label{sec:application-supporting-matrix}

The remaining studies help delimit the provisioning conclusions.
Tables~\ref{tab:application-fallback}--\ref{tab:application-repetition}
retain their explicit configurations and matched controls, including
cases where the endpoint changes little.

\paragraph{A component intervention can reverse its value.}
The configured GPU predecoder is slower than the CPU-only multiplier
reference at 32 CPU slots, under the reference mapping. At 16 slots, however,
the CPU-only reference takes 130.091\,ms and every tested predecoder
setting completes sooner, in 92.720--107.165\,ms
(Table~\ref{tab:application-fallback}). The reference against which
the component is evaluated changes the conclusion. Reducing summed
decoder work does not by itself establish a faster application: the
zero-fallback path uses 92.700\,ms of service but takes 92.720\,ms to
complete, whereas the 32-slot CPU-only path uses 1,129.275\,ms of
concurrent service and completes in 52.992\,ms. The configured service
graph and admission limits determine how that work becomes feedback.
The assumed predecoder profile does not support a measured GPU or
decoder-accuracy claim.

\begin{table*}[t]
\centering\small
\caption{Configured predecoder and CPU fallback. One factory chain and 24 outstanding windows. Three-seed medians in ms. M: multiplier. I: Ising. The predecoder uses one GPU slot and an assumed service of 5\% of the serial CPU window cost. These are timing sensitivities, not measured GPU performance or calibrated accuracy.}
\label{tab:application-fallback}
\begin{tabular}{@{}rlrrrrrr@{}}
\toprule
CPU & Fallback & M: time & M: service & M: QEC wait & I: time & I: service & I: QEC wait \\
\midrule
16 & CPU only & 130.091 & 2079.271 & 128.308 & 536.132 & 3437.081 & 347.968 \\
16 & 0\% & 92.720 & 92.700 & 90.441 & 536.052 & 171.841 & 63.400 \\
16 & 1\% & 95.362 & 114.387 & 93.180 & 536.052 & 206.244 & 86.743 \\
16 & 10\% & 106.594 & 317.730 & 104.815 & 536.118 & 514.864 & 225.464 \\
16 & 30\% & 107.165 & 749.566 & 105.583 & 536.127 & 1200.687 & 326.847 \\
32 & CPU only & 52.992 & 1129.275 & 47.057 & 536.132 & 3437.081 & 347.968 \\
32 & 0\% & 92.720 & 92.700 & 90.441 & 536.052 & 171.841 & 63.400 \\
32 & 1\% & 95.362 & 114.387 & 93.180 & 536.052 & 206.244 & 86.743 \\
32 & 10\% & 106.594 & 317.730 & 104.815 & 536.118 & 514.864 & 225.464 \\
32 & 30\% & 107.165 & 749.566 & 105.583 & 536.127 & 1200.687 & 326.847 \\
\bottomrule
\end{tabular}
\end{table*}

\paragraph{Storage matters when the supply regime makes it useful.}
The one-chain multiplier plateau beyond eight final-store entries is
visible in Table~\ref{tab:application-storage}. At four chains,
increasing capacity from four to 32 entries changes completion from
21.711 to 21.414\,ms and reduces accumulated material waiting from
2.667\,ms to zero. This is a modest endpoint improvement compared
with the CPU and factory interventions, but it locates a residual supply
constraint. Ising's endpoint remains unchanged at the displayed
precision throughout each store-capacity sweep. More inventory capacity
is therefore neither always useful nor always irrelevant.

\begin{table*}[t]
\centering\small
\caption{Final-state inventory capacity at 32 CPU slots. M: multiplier. I: Ising. Three-seed medians in ms. Material waiting is a sum over consumers and can exceed elapsed completion. Capacity limits do not equal actual peak occupancy.}
\label{tab:application-storage}
\begin{tabular}{@{}rrrrrr@{}}
\toprule
Chains & Store entries & M: time & M: material wait & I: time & I: material wait \\
\midrule
1 & 4 & 53.031 & 234.574 & 536.132 & 639.138 \\
1 & 8 & 52.992 & 234.574 & 536.132 & 639.138 \\
1 & 16 & 52.992 & 234.574 & 536.132 & 639.138 \\
1 & 32 & 52.992 & 234.574 & 536.132 & 639.138 \\
4 & 4 & 21.711 & 2.667 & 214.643 & 0.021 \\
4 & 8 & 21.605 & 1.671 & 214.643 & 0.021 \\
4 & 16 & 21.453 & 0.214 & 214.643 & 0.021 \\
4 & 32 & 21.414 & 0.000 & 214.643 & 0.021 \\
\bottomrule
\end{tabular}
\end{table*}

\paragraph{Control and factory-dependence effects are bounded here.}
Table~\ref{tab:application-controller} retains all three controller
service settings, including the 600\,ns reference.
Table~\ref{tab:application-dependence} extends the dependence comparison
to both one and four chains. The retained and constructed alternative
factory laws have identical median completion at the displayed precision,
while their accumulated material waiting differs slightly. These experiments bound
the effects in the evaluated configurations. They do not justify omitting
feedback application or outcome dependence from the execution model.

\begin{table*}[t]
\centering\small
\caption{Controller application service, one chain and 32 CPU slots. Median complete-application time across three seeds. All other reference settings are retained.}
\label{tab:application-controller}
\begin{tabular}{@{}lrr@{}}
\toprule
Controller service & Multiplier (ms) & Ising (ms) \\
\midrule
100 ns & 52.979 & 536.126 \\
600 ns & 52.992 & 536.132 \\
1000 ns & 52.997 & 536.137 \\
\bottomrule
\end{tabular}
\end{table*}

\begin{table*}[t]
\centering\small
\caption{Factory outcome/timing dependence under retained and constructed alternative laws, at 32 CPU slots. M: multiplier. I: Ising. Values are three-seed medians in ms. The unchanged completion times coexist with small differences in accumulated material waiting.}
\label{tab:application-dependence}
\begin{tabular}{@{}rlrrrr@{}}
\toprule
Chains & Law & M: time & M: material wait & I: time & I: material wait \\
\midrule
1 & retained & 52.992 & 234.574 & 536.132 & 639.138 \\
1 & independent & 52.992 & 234.415 & 536.132 & 637.275 \\
1 & correlated & 52.992 & 234.426 & 536.132 & 637.257 \\
4 & retained & 21.453 & 0.214 & 214.643 & 0.021 \\
4 & independent & 21.453 & 0.183 & 214.643 & 0.021 \\
4 & correlated & 21.453 & 0.172 & 214.643 & 0.021 \\
\bottomrule
\end{tabular}
\end{table*}

\paragraph{Repetition checks the complete execution boundary.}
The multiplier completes two repetitions in 106.155\,ms. Ising takes
1,072.515\,ms (Table~\ref{tab:application-repetition}).
The resulting per-repetition times are 53.077\,ms for the multiplier
and 536.258\,ms for Ising. Both remain close to their respective
single-repetition endpoints of 52.992 and 536.132\,ms.
Consumption doubles exactly to 1,400 and 14,220 T
states. Final drain remains a separate, later endpoint. This comparison
checks complete repeated work under retained state and scheduling rules. Two repetitions do not establish long-run steady-state throughput.

\begin{table*}[t]
\centering\small
\caption{Complete repeated execution with one chain and 32 CPU slots. Median times across three seeds. Drain includes work after application completion. Per-repetition time divides the whole job by its repetition count. T consumption is exact in all three seeds.}
\label{tab:application-repetition}
\begin{tabular}{@{}lrrrrr@{}}
\toprule
Application & Repetitions & Job (ms) & Per repetition (ms) & Drain (ms) & T states \\
\midrule
Multiplier & 1 & 52.992 & 52.992 & 54.064 & 700 \\
Multiplier & 2 & 106.155 & 53.077 & 107.153 & 1400 \\
Ising & 1 & 536.132 & 536.132 & 537.357 & 7110 \\
Ising & 2 & 1072.515 & 536.258 & 1073.888 & 14220 \\
\bottomrule
\end{tabular}
\end{table*}

\begin{table*}[t]
\centering\small
\caption{Coverage of the complete-application cohort across the main-text figures and supplemental tables. Each configuration has three seeds. Shared controls are reused without creating additional executions. The final column counts unique runs within each study. Controls may support more than one display.}
\label{tab:application-coverage}
\begin{tabular}{@{}lrrr@{}}
\toprule
Study & Configurations & Executions & In figures or tables \\
\midrule
Two-chain joint provisioning & 2 & 6 & 6 \\
CPU capacity and factory count & 24 & 72 & 72 \\
Decoder speed and factory count & 12 & 36 & 36 \\
Controller service & 4 & 12 & 12 \\
Complete repetition & 2 & 6 & 6 \\
Final-state storage & 12 & 36 & 36 \\
History and concurrency & 12 & 36 & 36 \\
Predecoder and fallback & 16 & 48 & 48 \\
Overlap & 8 & 24 & 24 \\
Reference service variability & 12 & 36 & 36 \\
Factory dependence & 8 & 24 & 24 \\
Family, size and supply & 60 & 180 & 180 \\
Family service variability & 8 & 24 & 24 \\
Family windows & 16 & 48 & 48 \\
Two-chain speed follow-up & 4 & 12 & 12 \\
Total & 200 & 600 & 600 \\
\bottomrule
\end{tabular}
\end{table*}

\paragraph{Cohort accounting.}
The main figures and these supporting tables retain the original 196
configuration classes and 588 executions. Figure~\ref{fig:application-provisioning-matrix}
adds four two-chain speed configurations with three seeds each, bringing
the total to 200 classes and 600 executions. Shared controls appear in several comparisons
but count once in the cohort. The later policy/resource campaign is a
separate evidence set.

% Flush supplemental floats before the class's final balancing hook.
\clearpage
\end{document}